\documentclass[longauth]{aa} 

\usepackage{graphicx, comment}
\usepackage{commath}
\usepackage{txfonts}
\usepackage{subfig}
\usepackage{upgreek}
\usepackage{threeparttable}
\usepackage{booktabs}      

\usepackage[colorlinks=true]{hyperref}

\hypersetup{
     colorlinks   = true,
     citecolor    = blue
}

\usepackage{orcidlink}

\newcommand{\C}{3C\,84}

\begin{document} 
\renewcommand{\arraystretch}{1.5}

    \title{Where within the 3C 84 jet are $\gamma$-rays produced?}
    \author{
Georgios~F. Paraschos\inst{1,2,3} \orcidlink{0000-0001-6757-3098},
Ioannis Liodakis\inst{4,3}\orcidlink{0000-0001-9200-4006},
Svetlana Jorstad\inst{5,6}\orcidlink{0000-0001-9522-5453},
Yuri Y. Kovalev\inst{3}\orcidlink{0000-0001-9303-3263},
Sudip Chakraborty\inst{7},
Fr\'{e}d\'{e}ric Marin\inst{8}\orcidlink{0000-0003-2049-2690},
Steven R. Ehlert\inst{7} \orcidlink{0000-0003-4420-2838},
Efthalia Traianou\inst{8,3}\orcidlink{0000-0002-1209-6500},
Lena C. Debbrecht\inst{3}\orcidlink{0009-0003-8342-4561},
Iv\'an Agudo\inst{10}\orcidlink{0000-0002-3777-6182},
Thibault Barnouin\inst{8}\orcidlink{0000-0003-1340-5675},
Jacob J. Casey\inst{11}\orcidlink{0009-0009-3051-6570},
Laura Di Gesu\inst{12}\orcidlink{0000-0002-5614-5028},
Philip Kaaret\inst{7}\orcidlink{0000-0002-3638-0637},
Dawoon E. Kim\inst{13}\orcidlink{0000-0001-5717-3736},
Fabian Kislat\inst{11}\orcidlink{0000-0001-7477-0380},
Ajay Ratheesh\inst{14,13}\orcidlink{0000-0003-0411-4243},
M. Lynne Saade\inst{15,7}\orcidlink{0000-0001-7163-7015},
Francesco Tombesi\inst{16,17}\orcidlink{0000-0002-6562-8654},
Alan Marscher\inst{5}\orcidlink{0000-0001-7396-3332},
Jos\'{e}-Luis G\'{o}mez\inst{10}\orcidlink{0000-0003-4190-7613},
Alexander B. Pushkarev\inst{18,19,20}\orcidlink{0000-0002-9702-2307}, 
Tuomas Savolainen\inst{21,2,3}\orcidlink{0000-0001-6214-1085}, 
Ioannis Myserlis\inst{22}\orcidlink{0000-0003-3025-9497},
Mark Gurwell\inst{23}\orcidlink{0000-0003-0685-3621},
Garrett Keating\inst{23}\orcidlink{0000-0002-3490-146X},
Ramprasad Rao\inst{23}\orcidlink{0000-0002-1407-7944},
Sincheol Kang\inst{24}\orcidlink{0000-0002-0112-4836},
Sang-Sung Lee\inst{24,25}\orcidlink{0000-0002-6269-594X},
Sanghyun Kim\inst{24}\orcidlink{0000-0001-7556-8504},
Whee Yeon Cheong\inst{24}\orcidlink{0009-0002-1871-5824},
Hyeon-Woo Jeong\inst{24,25}\orcidlink{0009-0005-7629-8450},
Chanwoo Song\inst{24,25}\orcidlink{0009-0003-8767-7080},
Shan Li\inst{24,25}\orcidlink{0009-0006-1247-0976},
Myeong-Seok Nam\inst{24,25}\orcidlink{0009-0001-4748-0211},
Diego \'{A}lvarez-Ortega\inst{4,26}\orcidlink{0000-0002-9998-5238},
Carolina Casadio\inst{4,26}\orcidlink{0000-0003-1117-2863},
Chien-Ting Chen\inst{15}\orcidlink{0000-0002-4945-5079},
Enrico Costa\inst{13}\orcidlink{0000-0003-4925-8523},
Eugene Churazov\inst{27}\orcidlink{0000-0002-0322-884X},
Riccardo Ferrazzoli\inst{13}\orcidlink{0000-0003-1074-8605},
Giorgio Galanti\inst{28}\orcidlink{0000-0001-7254-3029},
Ildar Khabibullin\inst{29,27}\orcidlink{0000-0003-3701-5882},
Stephen L. O'Dell\inst{7}\orcidlink{0000-0002-1868-8056},
Luigi Pacciani\inst{13}\orcidlink{0000-0001-6897-5996},
Marco Roncadelli\inst{30}\orcidlink{0009-0008-7799-584X},
Oliver J. Roberts\inst{15}\orcidlink{0000-0002-7150-9061},
Paolo Soffitta\inst{13}\orcidlink{0000-0002-7781-4104},
Douglas A. Swartz\inst{15}\orcidlink{0000-0002-2954-4461},
Fabrizio Tavecchio\inst{31}\orcidlink{0000-0003-0256-0995},
Martin C. Weisskopf\inst{7}\orcidlink{0000-0002-5270-4240},
Irina Zhuravleva\inst{32}\orcidlink{0000-0001-7630-8085}
}
    \authorrunning{G.~F. Paraschos et al.}
    \institute{
$^{1}$Finnish Centre for Astronomy with ESO, University of Turku, 20014 Turku, Finland\\
$^{2}$Aalto University Metsähovi Radio Observatory, Metsähovintie 114, FI-02540 Kylmälä, Finland\\
$^{}$\ \email{gfpara@utu.fi}\\
$^{3}$Max-Planck-Institut f\"{u}r Radioastronomie, Auf dem H\"{u}gel 69, D-53121 Bonn, Germany\\
$^{4}$Institute of Astrophysics, Foundation for Research and Technology - Hellas, Voutes, 7110, Heraklion, Greece\\
$^{5}$Institute for Astrophysical Research, Boston University, 725 Commonwealth Avenue, Boston, MA 02215, USA\\
$^{6}$Saint Petersburg State University, 7/9 Universitetskaya nab., St. Petersburg, 199034 Russia\\
$^{7}$NASA Marshall Space Flight Center, Huntsville, AL 35812, USA\\
$^{8}$Universit\'{e} de Strasbourg, CNRS, Observatoire Astronomique de Strasbourg, UMR 7550, 67000 Strasbourg, France\\
$^{9}$Interdisziplinäres Zentrum für wissenschaftliches Rechnen (IWR), Ruprecht-Karls-Universität Heidelberg, Im Neuenheimer Feld 205, 69120, Heidelberg, Germany\\
$^{10}$Instituto de Astrof\'{i}sica de Andaluc\'{i}a, IAA-CSIC, Glorieta de la Astronom\'{i}a s/n, E-18008 Granada, Spain\\
$^{11}$Department of Physics and Astronomy and Space Science Center, University of New Hampshire, Durham, NH 03824, USA\\
$^{12}$ASI - Agenzia Spaziale Italiana, Via del Politecnico snc, 00133 Roma, Italy\\
$^{13}$INAF, Istituto di Astrofisica e Planetologia Spaziali, Via Fosso del Cavaliere 100, 00133 Roma, Italy\\
$^{14}$Physical Research Laboratory, Thaltej, Ahmedabad, Gujarat 380009, India\\
$^{15}$Science \& Technology Institute, Universities Space Research Association, 320 Sparkman Drive, Huntsville, AL 35805, USA\\
$^{16}$Dipartimento di Fisica, Universit\'{a} degli Studi di Roma "Tor Vergata", Via della Ricerca Scientifica 1, 00133 Roma, Italy\\
$^{17}$Istituto Nazionale di Fisica Nucleare, Sezione di Roma "Tor Vergata", Via della Ricerca Scientifica 1, 00133 Roma, Italy\\
$^{18}$Crimean Astrophysical Observatory, 298409 Nauchny, Crimea\\
$^{19}$Institute for Nuclear Research of the Russian Academy of Sciences, 60th October Anniversary Prospect 7a, Moscow 117312, Russia\\
$^{20}$Lebedev Physical Institute, Pushchino Radio Astronomy Observatory, Radiotelescopnaya 1a, Pushchino 142290, Russia\\
$^{21}$Aalto University Department of Electronics and Nanoengineering, PL 15500, FI-00076 Aalto, Finland\\
$^{22}$Institut de Radioastronomie Millim\'{e}trique, Avenida Divina Pastora, 7, Local 20, E–18012 Granada, Spain\\
$^{23}$Center for Astrophysics $|$ Harvard \& Smithsonian, 60 Garden Street, Cambridge, MA 02138 USA\\
$^{24}$Korea Astronomy and Space Science Institute, 776 Daedeok-daero, Yuseong-gu, Daejeon 34055, Korea\\
$^{25}$University of Science and Technology, Korea, 217 Gajeong-ro, Yuseong-gu, Daejeon 34113, Korea\\
$^{26}$Department of Physics, University of Crete, 70013, Heraklion, Greece\\
$^{27}$Max Planck Institute for Astrophysics, Karl-Schwarzschild-Str. 1, D-85741 Garching, Germany\\
$^{28}$INAF, Istituto di Astrofisica Spaziale e Fisica Cosmica di Milano, Via Alfonso Corti 12, I – 20133 Milano, Italy\\
$^{29}$Universit\"{a}ts-Sternwarte, Fakult\"{a}t für Physik, Ludwig-Maximilians-Universit\"{a}t München, Scheinerstr.1, 81679 München, Germany\\
$^{30}$INFN, Sezione di Pavia, Via A. Bassi 6, 27100 Pavia, Italy\\
$^{31}$INAF Osservatorio Astronomico di Brera, Via E. Bianchi 46, 23807 Merate (LC), Italy\\
$^{32}$Department of Astronomy and Astrophysics, The University of Chicago, Chicago, IL 60637, USA\\
}

   \date{Received -; accepted -}

 \abstract{
   The location of $\gamma$-ray creation and emission within extra-galactic jets is a matter of active debate.
   One particularly well-suited source to pinpoint the location is the nearby, bright radio galaxy 3C\,84, harbouring a powerful jet.
   Here we investigate the origin of $\gamma$-rays measured during a recent $\gamma$-ray flare, by analysing the linear polarisation signal of close-in-time very long baseline interferometry (VLBI) observations at centimetre and millimetre wavelengths.
   While 3C\,84 is overall almost unpolarised, we find that close-in-time to the $\gamma$-ray flare peak regions at parsec-scale distances from the central engine shows a fractional linear polarisation increase.
   Under the physically well-motivated assumption of a causal relation between this polarisation enhancement and the $\gamma$-ray flare, and combined with insights from concurrent X-ray polarisation measurements, the $\gamma$-rays being created in this region is a physically motivated scenario, in a process consistent with synchrotron self-Compton.
   }

   \keywords{
            Galaxies: jets -- Galaxies: active -- Galaxies: individual: 3C\,84 (NGC\,1275) -- Techniques: interferometric -- Techniques: high angular resolution -- Techniques: polarimetric
               }

   \maketitle

\section{Introduction}

Flares bright in $\gamma$-rays are a common feature of the light-curves of astrophysical jets.
It is thought that such flares are a direct result of the highly energetic beam of plasma powered by a central supermassive black hole (SMBH) inside an active galactic nucleus (AGN) interacting with itself and its environment \citep[see e.g.][]{MacDonald17, Lico17}.
However, questions still remain about the exact location of their origin within the entirety of the jet.
Trying to answer these questions requires investigating powerful nearby jets, where we can utilise the resolving ability of centimetre and millimetre very long baseline interferometry (VLBI) to directly probe the possible emission regions with high fidelity.

For the work presented here we chose the nearby radio galaxy \C\ (NGC\,1275; $z=0.0176$, \citealt{Strauss92}), which harbours a powerful radio jet and also recently exhibited a bright $\gamma$-ray flare.
While \C\ has been studied for many decades \citep[e.g.][]{Backer87, Krichbaum92, Homan04, Nagai12, Paraschos22} in which  the source has fluctuated between periods of intense activity and comparative quiescence, it is still unclear where the $\gamma$-ray location stems from.
Attempts have been made in the past to pinpoint the exact location, by means of variability light curve correlation \citep[e.g.][]{Hodgson18, Hodgson21, Paraschos23, Sinitsyna25}, as well as by investigating structural changes within the jet \citep{Abdo09, Nagai10}.
The former studies came to the conclusion that the long-term variability is consistent with multiple emission regions, both from near the jet origin and from further downstream in the parsec scale region.
The latter studies, which investigated the connection between a radio and $\gamma$-ray flare in 2008, favour a downstream emission region.

In our study we explore the connection between a recent $\gamma$-ray flare reported in \C\ and the concurrent change in linear polarisation within the jet.
This work is motivated by the multi-wavelength campaign that was initiated by the Imaging X-ray Polarimetry Explorer \citep[IXPE;][]{Weisskopf2022, Soffitta23}, with an observing time of the source between January and March of 2025.
These observations are the longest IXPE observations of a source ever to-date, lasting 2.5\,Msec \citep[see][for details]{Liodakis25}.
\C\ is known for exhibiting an ordered magnetic field associated with its jet \citep[see][]{Paraschos24b, Paraschos24a} and an increase in linear polarisation is a signature of the magnetic field's enhancement \citep[see also][]{MacDonald21, Kramer24}, which can be connected to flaring activity.
We note, however, that \C\ generally exhibits low amounts of linear polarisation, of the order of a percent at 43\,GHz, but increasing with frequency \citep[see, e.g.][]{Nagai17, Kim19}.
Here we report an enhancement of the linear polarisation at a distance of $\sim1.5$\,parsec\footnote{For our work we assume a $\Lambda$ cold dark matter cosmology ($\Lambda$CDM) model, with $H_0 = 67.8 \textrm{kms}^{-1} \textrm{Mpc}^{-1}$, $\Omega_\Lambda = 0.692$, and $\Omega_\textrm{M} = 0.308$ \citep{Planck16}, which results in a megaparsec luminosity distance of $D_\textrm{L} = 78.9 \pm2.4\,\textrm{Mpc}$ and a parsec-to-milliarcsecond (pc-mas) conversion factor of $\psi = 0.36\,\textrm{pc mas}^{-1}$.} from the central engine \citep{Paraschos21} during the 2025 $\gamma$-ray flare.
In Sect.~\ref{sec:Results} we present the observations and our results, which we discuss in Sect.~\ref{sec:Discussion}.
Finally, in Sect.~\ref{sec:Conclusions} we summarise our work.

\section{Methods}\label{sec:Results}

\subsection{Observations}

\C\ was observed in VLBI mode at 15\,GHz and 43\,GHz; specifically, the 15\,GHz VLBA sessions were observed as part of the MOJAVE programme \citep{Lister18}\footnote{\url{https://www.cv.nrao.edu/MOJAVE/index.html}} and the 43\,GHz VLBA sessions were conducted as part of the BEAM-ME programme \citep{Jorstad16}\footnote{\url{https://www.bu.edu/blazars/BEAM-ME.html}}.
The epochs we used are presented in Table~\ref{tab:Epochs}.
Furthermore, \C\ was observed in single-dish and connected interferometer (s.d./c.i) mode between 22\,GHz and 225\,GHz.
These observations were conducted between mid of January 2025 and beginning of April 2025 (see Fig.~\ref{fig:SingleDish}).
The participating telescopes were the Korean VLBI Network \citep[KVN;][observing at 22, 43, 86, and 129\,GHz]{Kang15}, the IRAM-30m telescope \citep[via the Polarimetric Monitoring of AGN at Millimeter Wavelengths programme][observing at 86 and 225\,GHz]{Agudo18a}, and the Submillimeter Array (SMA; via the SMA Monitoring of AGNs with Polarization, see Myserlis et al. in prep.).
Relevant details are discussed in \cite{Liodakis25}.

In a fortunate turn of events, \cite{Guarnieri25} reported in early January of 2025 the enhancement of $\gamma$-ray emission from a location consistent with \C\ as detected with the \textit{Fermi} Large Area Telescope \citep[LAT;][]{Abdollahi23}, with activity spanning between December 2024 and March 2025.
These $\gamma$-ray observations, pulled directly from the publicly available repository\footnote{\url{https://fermi.gsfc.nasa.gov/ssc/data/access/lat/LightCurveRepository/}}, binned in three day intervals using a free spectral index, are also presented in Fig.~\ref{fig:SingleDish}.
As this time frame coincides with our VLBI measurements, we were able to investigate the changes in the jet before, during, and after this flare.

\subsection{Data analysis}

We utilised all available during the time frame of interest epochs from the MOJAVE and BEAM-ME monitoring programmes, which provide publicly available, calibrated VLBI data \citep[see][for a description of the calibration and imaging procedures]{Jorstad17, Lister18}.
We re-imaged the BEAM-ME data using the regularised maximum likelihood (RML) method implemented in the \texttt{eht-imaging} software suite \citep{Chael16, Chael18}, and used the MOJAVE data as they are provided by the MOJAVE collaboration, without reanalysis.
Given the high brightness of \C\ at 43\,GHz, we achieved high-resolution imaging that recovered detailed total intensity and polarimetric structure.
Unlike traditional \texttt{CLEAN}-based approaches, RML uses forward modelling with physically motivated regularisers -- such as total variation, entropy, and sparsity -- to manage sparse VLBI coverage while preserving fine-scale structure \citep{EHT19a, EHT22a}. 
This approach is particularly well-suited for complex sources like \C\ characterised by filamentary structures, for which geometrical model fitting may be less reliable \citep[see, e.g.][for an implementation on core-jet morphologies]{Paraschos24c}.
Furthermore, the possibility of super-resolution compared to \texttt{CLEAN} that RML offers facilitates the exact spatial pinpointing of the polarised emission.

In order to determine the regularisation terms needed to best fit all data sets, we varied them across a search grid, after having accounted for non-closing systematic errors of 1\% of visibility amplitude \citep[see][for more details on the procedure, which involves solving for both complex visibilities and closure quantities, and the regularisation terms]{Traianou2025}.
These terms include the $\ell1$ norm, the relative entropy $\textrm{mem}$, the total variation tv, the total squared variation tv2, and the total variation $\ell2$ with the logarithmic regularising term tv2log.
The set of regularisation terms fitting all epochs the best ($\chi^2\approx1$) were: $\ell1=0.6$, $\textrm{mem}=0.1$, $\textrm{tv}=0$, $\textrm{tv2}=1.3$, and $\textrm{tv2log}=0.02$, for the corresponding data terms $\alpha_\textrm{amp}=50$, $\alpha_\textrm{cp}=90$, and $\alpha_\textrm{lca}=90$ \citep[see also][]{Janssen21}.

On the other hand, while the 15\,GHz observations offer higher sensitivity, the source is known to exhibit a decreasing linear polarisation flux density trend with decreasing frequency.
Therefore, we do not expect impactful data for our present analysis, as also shown in Sect.~\ref{sec:Results}.
As a sanity check, we also compared our results with the publicly available \texttt{CLEAN} images of BEAM-ME.
By employing both methods of reconstruction, performed by different teams (given that the publicly available nature of the 43\,GHz data were analysed independently by the BEAM-ME team), we achieve added robustness to our results.
We found good agreement between the overall jet structure (although resolved in more detail in our approach) and flux densities (see also Fig.~\ref{fig:BU2} in Appendix~\ref{app:BU}).

Importantly for this work, \texttt{eht-imaging} simultaneously reconstructs Stokes Q and U during polarimetric imaging, applying constraints such as the Holdaway-Wardle limit and polarimetric total variation to ensure physical consistency \citep{Holdaway1990, EHT21a, Traianou2025}. 
The integration of instrumental calibration within the imaging loop further improves convergence and image fidelity, making the method well suited to low-polarisation sources like \C\ \citep{EHT21a}.

\begin{table*}
\centering
\begin{threeparttable}

\caption{VLBI epochs used in this work.}\label{tab:Epochs}

\begin{tabular}{c|c}
  \multicolumn{1}{c}{BEAM-ME (43\,GHz)} & \multicolumn{1}{c}{MOJAVE (15\,GHz)} \\[0.3em]
  \begin{tabular}{ccccc}
    \hline\hline
    Date & $\textrm{I}_\textrm{Core}$\,[Jy] & $\textrm{PD}_\textrm{Core}$ & $\textrm{I}_\textrm{C3}$\,[Jy] & $\textrm{PD}_\textrm{C3}$ [\%]\\
    \hline
    04 Apr. 2025 & 6.7$\pm$0.7  & <0.7         & 0.7$\pm$0.1 & 7$\pm$2.1  \\
    23 Mar. 2025 & 8.5$\pm$0.9  & 1.1$\pm$0.2  & 0.9$\pm$0.1 & 7$\pm$1.7  \\
    09 Mar. 2025 & 8.3$\pm$0.8  & 1.3$\pm$0.2  & 0.7$\pm$0.1 & <7.7       \\
    16 Feb. 2025 & 7.7$\pm$0.8  & 1.2$\pm$0.2  & 0.8$\pm$0.1 & 9$\pm$2.0  \\
    15 Dec. 2024 & 8.6$\pm$0.9  & 1.2$\pm$0.2  & 1.1$\pm$0.1 & 12$\pm$1.8 \\
    21 Nov. 2024 & 10.0$\pm$1.0 & 1.0$\pm$0.2  & 1.0$\pm$0.1 & 5$\pm$1.4  \\
    \hline
  \end{tabular}
  &
  \begin{tabular}{ccc}
    \hline\hline
    Date & $\textrm{I}_\textrm{tot}$\,[Jy] & $\textrm{PD}_\textrm{tot}$ [\%]\\
    \hline
    16 Mar. 2025 & 32.7 $\pm$ 1.6 & $\sim$0.1 \\
    21 Feb. 2025 & 45.4 $\pm$ 2.3 & <0.1      \\
    26 Jan. 2025 & 40.7 $\pm$ 2.0 & $\sim$0.1 \\
    06 Jan. 2025 & 45.1 $\pm$ 2.3 & $\sim$0.1 \\
    26 Nov. 2024 & 42.5 $\pm$ 2.1 & <0.1      \\
    & & \\
    \hline
  \end{tabular}
\end{tabular}

\begin{tablenotes}
\footnotesize
\item The left part of the table corresponds to the 43\,GHz BEAM-ME observations and the right part to the 15\,GHz MOJAVE (\texttt{CLEAN}-reconstructed) observations.
From left to right the columns indicate the observation date, the total intensity of the area of interest (Core and C3 for the 43\,GHz observations, total for the 15\,GHz ones), and the linear polarisation degree.
They are calculated by integrating the flux density within the area the size of the nominal \texttt{CLEAN} beam.
\end{tablenotes}

\end{threeparttable}
\end{table*}

\subsection{Results}

\begin{figure*}
\centering
\includegraphics[scale=0.7]{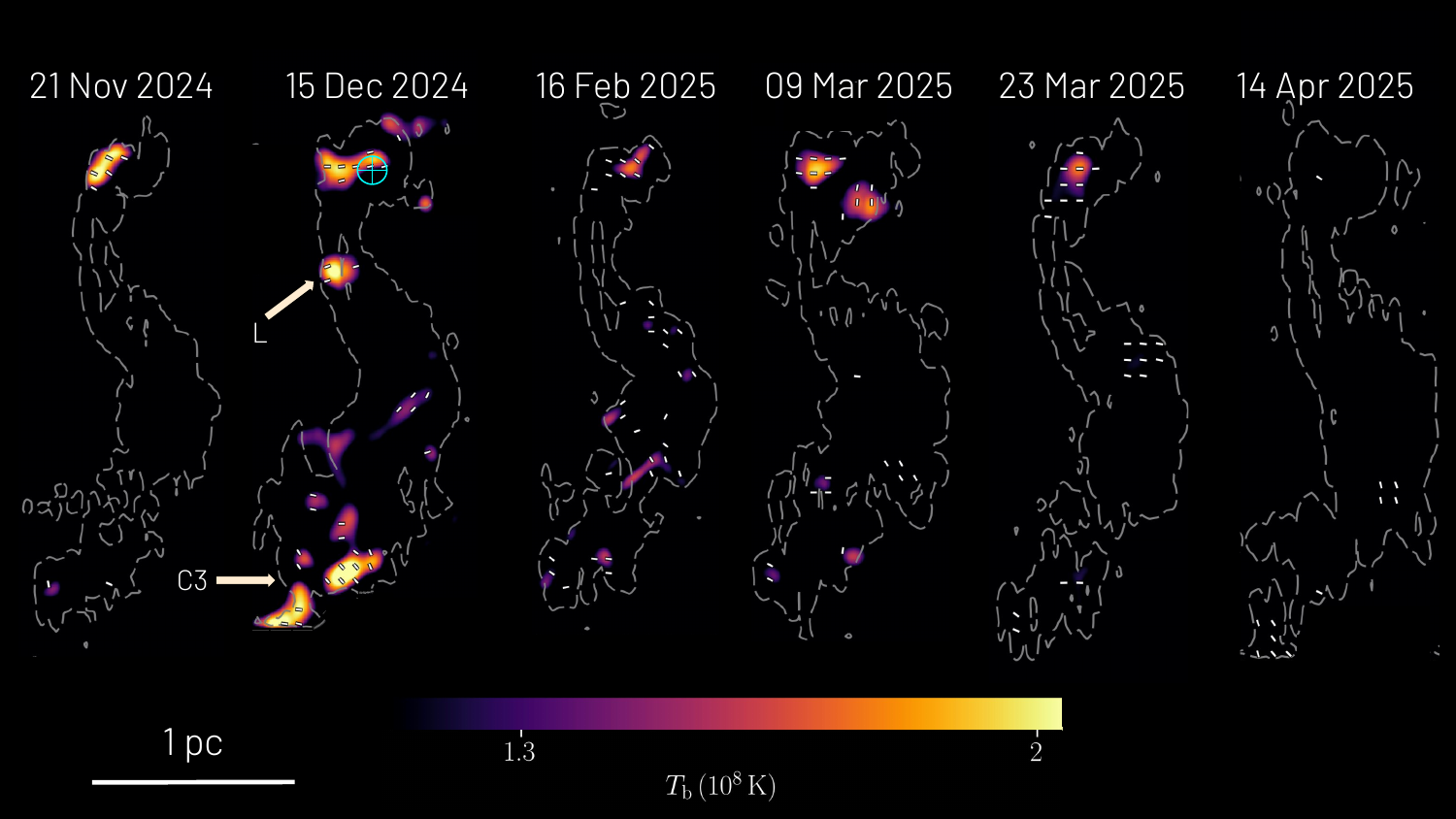}
\includegraphics[scale=0.7]{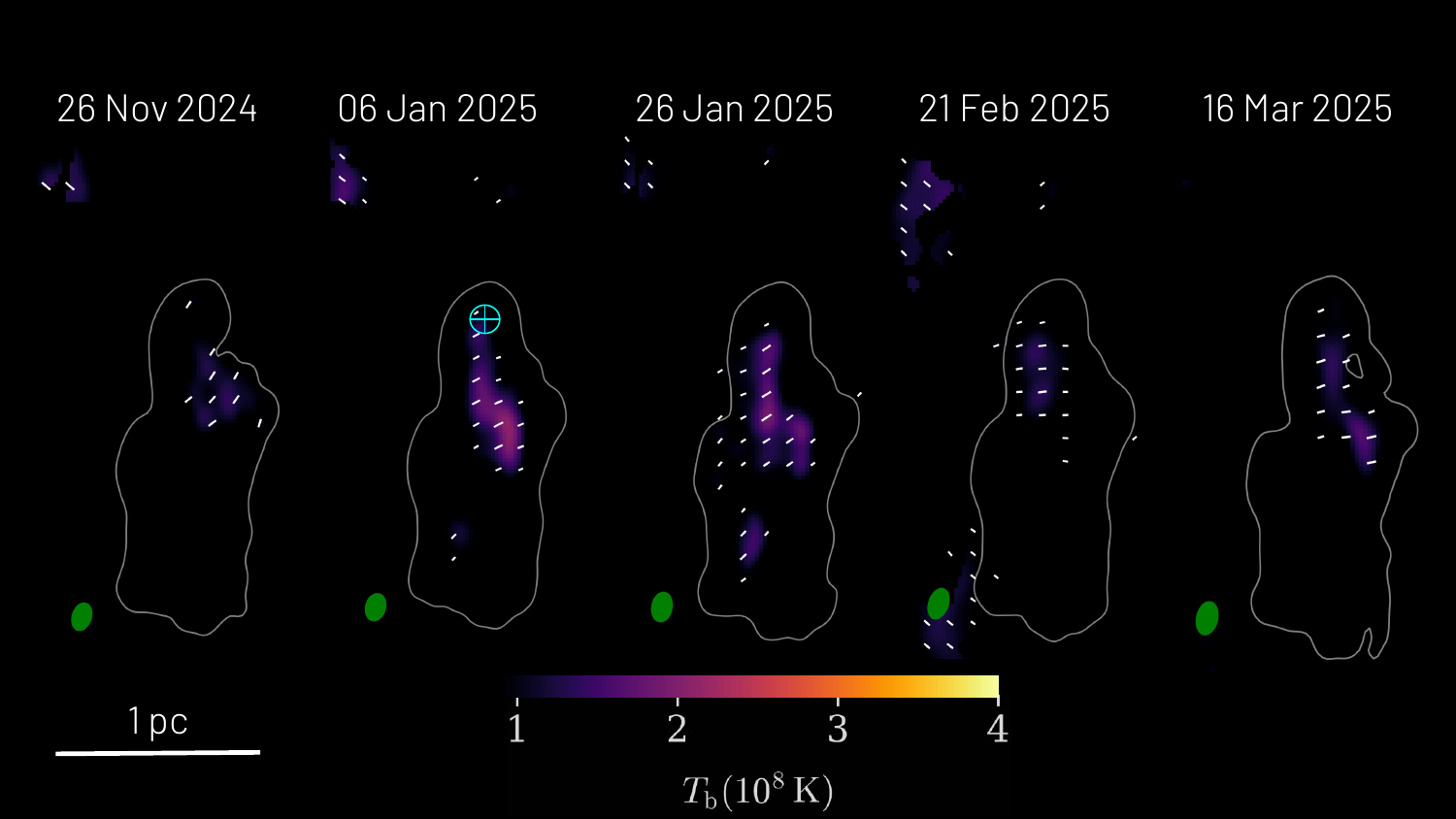}
  \caption{
    Images of \C\ at 43\,GHz (BEAM-ME) and 15\,GHz (MOJAVE).
    \emph{Top panel}: The 43\,GHz total intensity (contours) and linear polarisation (colour) images of \C\ for all available epochs during the time frame of interest are shown here.
    The contour levels were set at 1\% of the total intensity maximum ($I_\textrm{max}$) per epoch. 
    The linear polarisation flux density is displayed in units of brightness temperature.
    The chosen colour scale is meant to display the detections, while also masking the surrounding noise.
    The cyan cross and circle denote the core region and the arrows C3, as discussed in the main text, and region L (see discussion in Sect.~\ref{sec:Discussion}).
    The white sticks indicate the EVPAs.
    At the epoch nearest to the $\gamma$-ray flare peak (15 Dec. 2024) we detect linear polarisation at a distance of $\sim1.5$\,parsec (C3) from the central engine.
    \emph{Bottom panel}: The 15\,GHz images of the \C\ jet for the available epochs in the time frame of interest, shown in a similar manner as in the top panel, prepared by the MOJAVE collaboration (\texttt{CLEAN}-reconstruction).
    The green ellipse next to each reconstruction illustrates the restoring \textrm{CLEAN}-beam, which corresponds to $(0.8\times0.5)\,\textrm{mas}\,(15^\circ)$ on average.
    We note that the linear polarisation is similar in magnitude to the noise level, amounting to marginal/non-detections.
    }
   \label{fig:BU}
\end{figure*}

Here we present the results of our millimetre VLBI analysis, shown in Fig.~\ref{fig:BU} (top panel) and also discuss them below (our centimetre VLBI analysis, exhibiting a lower overall linear polarisation signal consistent with the source's history is shown in bottom panel; total intensity images are available online).
In both instances the images are aligned on the so-called radio core.
Defining and identifying the core in a VLBI image is a delicate matter, subject to a number of assumptions.
In our case, we follow the vast number of publications available for \C\ over the years \citep[see, e.g.][for two recent examples of VLBI monitoring of the source at cm VLBI wavelengths over decades]{Lister18, Weaver22}, which place the centimetre VLBI core at the brightest area inside the northernmost part of the jet.
To aid the reader, we have denoted the core with a cyan cross and circle in both panels.
We find that \C\ is characterised by overall low total integrated polarisation, of the order of $\sim1\%$.
In the core we found a linear polarisation detection in most epochs (see Table~\ref{tab:Epochs}).
However, during the $\gamma$-ray flare peak (at the December 15 2024 epoch) we find a fractional linear polarisation enhancement, in regions downstream of the central engine.
Therefore, we focus on the region downstream of the core in the remainder of this work.
Specifically, the area 1.5\,parsec away from it (denoted as `C3'), identified with the jet termination region \citep{Kam24} is the one exhibiting a linear polarisation brightening.
The respective values are listed in Table~\ref{tab:Epochs}.
We point out here that the linear polarisation calibration uncertainty (commonly referred to as D-term calibration) is of the order of 0.9\%, indicating that our detections are robust.
This value was calculated by averaging the D-term corrections of all sources of the epochs in question, as calculated using the procedure described in \cite{Jorstad17}, given that we did not re-calibrate the already calibrated D-terms in our imaging procedure.

Furthermore, we note that a region (denoted with `L' in Fig.~\ref{fig:BU}) which is also outside the compact core, at a distance of $\sim0.5$\,parsec, exhibits a brightness increase as well, but only in the December 15th 2024 epoch.
For the MOJAVE measurements we find that the linear polarisation signal is weaker, as is consistent with the historical linear polarisation values of \C.

\begin{figure*}
\centering
\includegraphics[scale=0.45]{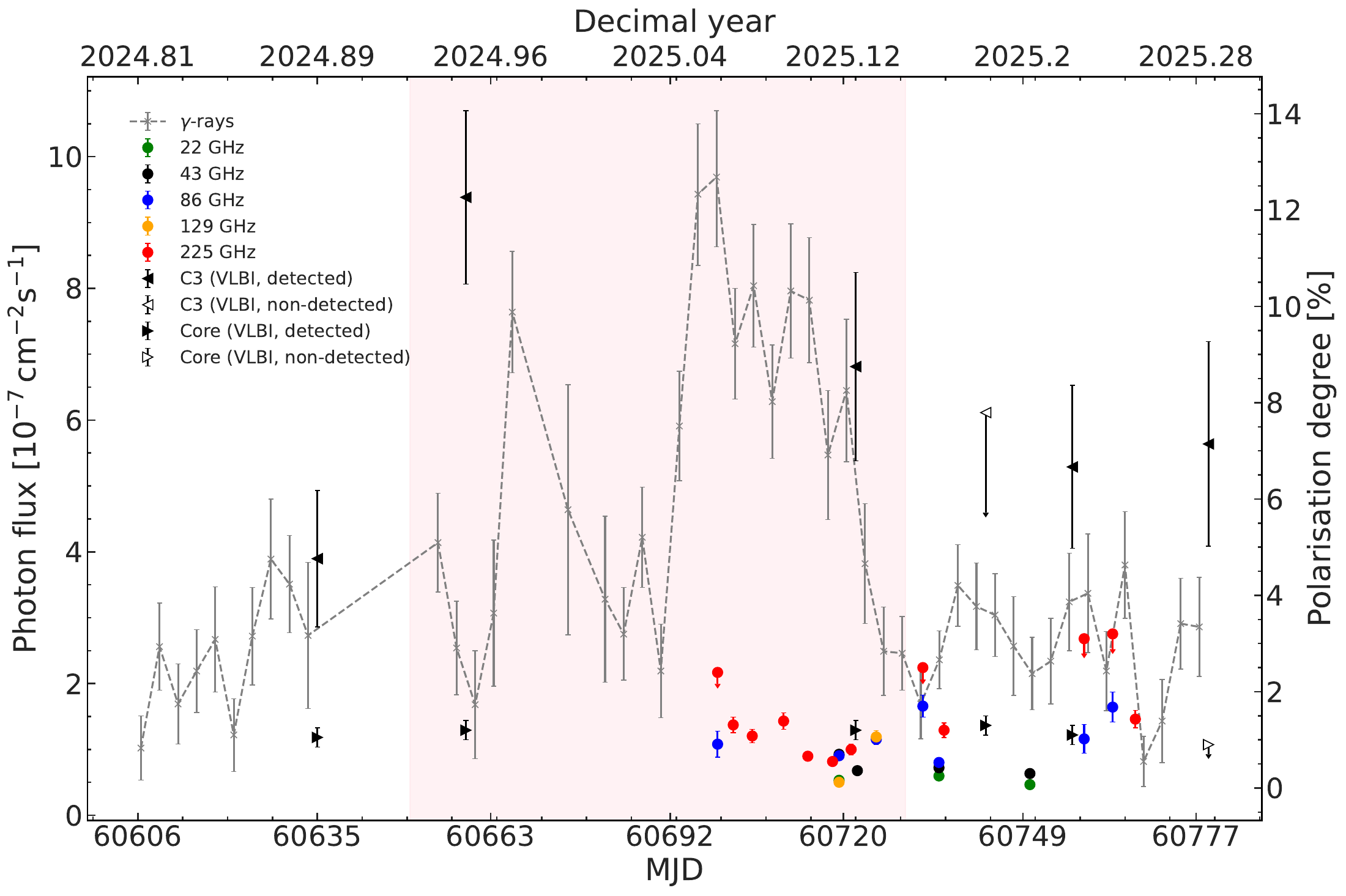}
  \caption{
    Linear polarisation degree and $\gamma$-ray flux as a function of time.
    Shown here are observations taken with the KVN at 22, 43, 86, and 129\,GHz, with the IRAM 30\,m telescope as part of the POLAMI programme at 86 and 225\,GHz, with the SMA at 225\,GHz (all in colour and round markers) and Fermi-LAT in $\gamma$-rays (grey crosses connected by a dashed line).
    In addition, the black triangular markers denote the VLBI polarisation degree of the core and C3 at 43\,GHz.
    The filled ones with an error bar correspond to a detection (close to the $\gamma$-ray peak, with the entirety of the flare being denoted with the light pink shaded area), the empty one to a non-detection upper limit.
    All values are listed in Table~\ref{tab:Epochs}.
    The well-known trend for \C\ of higher polarisation values at higher frequencies is observed here as well.
    The radio observation dates span from MJD 60696 to 60767.
    During this time-frame some depolarisation is also observed, possibly due to multiple jet components blurring the s.d./c.i measurements.
    }
   \label{fig:SingleDish}
\end{figure*}

Overall, as shown in Fig.~\ref{fig:SingleDish}, the VLBI polarisation degree exhibits higher values at the onset of the $\gamma$-ray flare, and then decreases in the aftermath (see also Fig.~\ref{fig:SingleDishSI} in Appendix~\ref{app:LC}, for the total intensity behaviour).
While the uncertainties are large, a clear trend exists, which is suggestive of this rise during the $\gamma$-ray flare being real.
A $\chi^2$ test against the hypothesis of a constant polarisation degree (in essence a model in which the ratio between Stokes I and the linear polarisation P remains constant per measurement and any deviation is attributed to measurement noise) yields $\chi^2=14.5$ for four degrees of freedom ($\chi^2_\textrm{red}=2.89$, $p=0.01$), indicating statistically significant variability during the observing window.
The s.d./c.i polarisation degree measurements hint at an enhancement after the $\gamma$-ray flare peak before dimming again.
They also follow the trend of higher linear polarisation at higher frequencies.
We note that the apparent depolarisation might be connected to the fact that the s.d./c.i observations capture the entirety of the jet, which can have a smearing effect, due to the different regions probed contributing to the measurements.
The electric vector position angles (EVPAs; see Table~\ref{tab:SingleDish}) at 22 and 225\,GHz appear aligned with the North-South jet direction, whereas the other frequencies seem to diverge.

\begin{table*}
    \centering
    \begin{threeparttable}
    \caption{Linear polarisation measurements of \C\ in s.d./c.i mode.}\label{tab:SingleDish}
    \begin{tabular}{cccc}
    \hline
    \hline
    Frequency [GHz] & MJD  & Polarisation degree [\%] & Position Angle [deg] \\
    \hline
         
        22$^*$       & 60719.3 & $0.16\pm0.01$ &  $-4.33\pm12.33$ \\
        22$^*$       & 60735.4 & $0.25\pm0.03$ &  $16.92\pm9.35$  \\
        22$^*$       & 60750.1 & $0.07\pm0.04$ &  $18.47\pm1.74$  \\
        43$^*$       & 60719.3 & $0.70\pm0.03$ &  $112.20\pm1.45$ \\
        43$^*$       & 60722.3 & $0.36\pm0.04$ &  $124.38\pm3.83$ \\
        43$^*$       & 60735.4 & $0.42\pm0.03$ &  $105.07\pm3.89$ \\
        43$^*$       & 60750.1 & $0.30\pm0.06$ &  $134.19\pm5.24$ \\
        86$^\dagger$ & 60699.7 & $0.91\pm0.27$ &  $51.5\pm7.0$    \\
        86$^*$       & 60722.3 & $0.67\pm0.10$ &  $76.60\pm3.20$  \\
        86$^*$       & 60725.3 & $1.01\pm0.10$ &  $49.93\pm3.59$  \\
        86$^\dagger$ & 60732.8 & $1.70\pm0.23$ &  $23.6\pm3.8$    \\
        86$^*$       & 60735.4 & $0.53\pm0.07$ &  $63.59\pm4.51$  \\
        86$^\dagger$ & 60758.9 & $1.02\pm0.30$ &  $24.8\pm7.7$    \\
        86$^\dagger$ & 60763.5 & $1.68\pm0.31$ &  $54.8\pm4.6$    \\
        129$^*$      & 60719.3 & $0.12\pm0.08$ &  $63.46\pm2.73$  \\
        129$^*$      & 60725.3 & $1.06\pm0.13$ &  $31.28\pm4.25$  \\
        225$^\dagger$ & 60699.7 & $<2.4$        &  -- \\
        225$^\Box$   & 60702.2 & $1.31\pm0.16$ &  $-28.80\pm1.85$ \\
        225$^\Box$   & 60705.3 & $1.08\pm0.14$ &  $-15.72\pm1.92$ \\
        225$^\Box$   & 60710.3 & $1.39\pm0.17$ &  $-26.21\pm1.98$ \\
        225$^\Box$   & 60714.3 & $0.66\pm0.09$ &  $-11.25\pm1.45$ \\
        225$^\Box$   & 60718.2 & $0.55\pm0.08$ &  $-7.07\pm1.01$ \\
        225$^\Box$   & 60721.3 & $0.80\pm0.11$ &  $-11.29\pm1.46$ \\
        225$^\dagger$ & 60732.8 & $<2.5$        &  -- \\
        225$^\Box$   & 60736.3 & $1.20\pm0.15$ &  $23.00\pm2.04$ \\
        225$^\dagger$ & 60758.9 & $<3.1$        &  -- \\
        225$^\Box$   & 60767.2 & $1.43\pm0.18$ &  $17.74\pm1.92$ \\
        225$^\dagger$ & 60763.5 & $<3.2$        &  -- \\
    \hline
    \end{tabular}

    \begin{tablenotes}
        \footnotesize
        \item $^*$KVN; $^\dagger$POLAMI; $^\Box$SMA
    \end{tablenotes}

    \end{threeparttable}
\end{table*}

\section{Discussion}\label{sec:Discussion}

Locating the exact position of $\gamma$-ray production within jets has been a topic of active research over recent years and nearby jetted radio galaxies offer a unique laboratory to directly examine the possible locations.
\C\ is such a jet, whose morphology has carefully been monitored over the years.
Its parsec scale jet consists of three characteristic regions, referred to as `C1' (core), `C2' (diffuse parsec scale region), and C3 (bright moving knot), connected by a double rail structure and surrounded by a cocoon \citep{Savolainen23}.
This filamentary structure, which reaches deep into the core region \citep{Nagai14, Giovannini18} and points to transverse jet stratification, has been interpreted as a possible manifestation of a Kelvin-Helmholtz instability \citep{Paraschos25b}.

In this complex structure pinpointing the exact location of the $\gamma$-ray emission zone is not straightforward.
Finding this location is particularly important, because it is thought to coincide with the elusive blazar zone \citep{Hovatta19b}, where particles are accelerated to the high energies we commonly observe in blazar jets.
Two scenarios are usually invoked: the near site \citep[within the broad line region which provides the photon field for upscattering, see][among others]{Sikora94, Blandford95} and the far site \citep[beyond the broad line region, e.g.][]{Lahteenmaki03, Lindfors06, Marscher12, Jorstad13, Kramarenko22}.
Past studies, which have tried to pinpoint the location of $\gamma$-ray emission within the \C\ jet \citep[e.g.][]{Hodgson18, Hodgson21, Paraschos23, Sinitsyna25} by investigating the long term behaviour of the multi-band light-curves of the source, have provided evidence in favour of both scenarios. 

In this work we found indications that $\gamma$-rays emitted during the recent, early 2025 flare originate far from the central engine, at a distance of $\sim1$\,parsec.
We base this assessment on our linear polarisation analysis, that revealed a brightness enhancement in that region, which is a signature for an ordered magnetic field amplitude increase.
Under the assumption that the linear polarisation enhancement is associated with the $\gamma$-ray flare, our result suggests that \C\ has likely been producing $\gamma$-rays downstream of the central engine for at least the last two decades \citep{Abdo09, Nagai10}.
This assumption is well motivated, as similar behaviour has been explained theoretically \citep[e.g.][]{Hughes11}, as well as noted for a number of sources in the literature, for example, OJ\,287 \citep{Agudo11}, 3C\,279 \citep{Rani17}, 3C\,454.3 \citep{Liodakis20}, OJ\,248 \citep{Paraschos25a}, PKS\,0735+178 \citep{Paraschos25c}, and 3C\,120 \citep[][]{Traianou26}.
While the examples above do not disprove a chance coincidence of a radio polarisation degree enhancement alongside a $\gamma$-ray flare, they do indicate that a causal connection is plausible.
Nevertheless, in these cases a transient increase of the fractional linear polarisation signals a shock or reconnection driven ordering and compression of the magnetic field. 
The ordered field raises the synchrotron emissivity and the local synchrotron photon energy density, which, in turn, produces the observed $\gamma$‑ray flare (possibly through synchrotron self-Compton, as discussed also further below).
We note that the brightness enhancement of region L only appears in the December 15th 2024 epoch; its absence from all other epochs, along with it being rather compact and localised, indicates that it could possibly be an imaging artefact due to low S/N in that area.
Stacking in time the total intensity and polarised BU images used here does not reveal any pronounced activity in that region.
If, however, this region is not an artefact then it is at a distance of $\sim1$ parsec to C3, indicating that they are most likely not causally connected; instead multiple emission regions, such as those created by multiple shocks, turbulence, or magnetic reconnection, for example, by jet-in-jet models, appear more likely \citep[see e.g.][]{Giannios09, Narayan12b, Clausen-Brown12, Biteau12, Dotson15}.

Furthermore, the accompanying to this work IXPE study by \cite{Liodakis25} revealed that the X-ray position angle is parallel to bulk jet flow and the X-ray polarisation is of the order of $4\%$.
The authors also showed that optical and $\gamma$-ray variability strongly correlates, at a time lag consistent with zero days, with the optical photons originating mainly from synchrotron emission, as a precursor of synchrotron self-Comptonised X-ray and $\gamma$-ray photons \citep{Marin25}.
They interpret their results as being caused by the synchrotron self-Compton (SSC) emission mechanism.

Motivated by these findings, we designed our analysis to address the plausibility of SSC in our case.
Our approach was two‑fold: we first performed a comparative analysis to ascertain whether or not the physical conditions present in the C3 region of \C\ are able to give rise to the observed $\gamma$-ray flare, whose peak value is of the order of $F^{\textit{Fermi}}_{\textrm{SSC}} \sim 4\times10^{-4}\,\textrm{MeV} /\textrm{cm}^{2}/ \textrm{s}$.
Assuming equipartition between the particle and magnetic energy, we started by estimating the magnetic field strength in the emitting region from the synchrotron self-absorption (SSA) turnover parameters following the procedure outlined in Appendix~D of \cite{Paraschos24a}, based on the formalism of \cite{Marscher83}. 
For a homogeneous, spherical synchrotron source of angular diameter $\theta_{\textrm{mas}}$ and turnover frequency $\nu_m$ [GHz] at which the observed flux density is $S^\prime_m$ [Jy], the SSA magnetic field in Gauss is given by:
\begin{equation}
B_{\textrm{SSA}} = 10^{-5} \; b(\alpha) \;
\theta_{\textrm{mas}}^{4} \;
\nu_{m}^{5} \;
S_m^{\prime-2} \;
\frac{\delta}{1+z} \, \;[\textrm{G}],
\label{eq:bssa}
\end{equation}
where $b(\alpha)$ is a dimensionless function tabulated by \citet{Marscher83} that depends on the optically thin spectral index $\alpha$ ($S_\nu \propto \nu^{\alpha}$), $\delta$ is the Doppler factor, and $z$ is the source redshift.
We adopted $\alpha = -0.5$, for which $b(\alpha) \simeq 3.2$, set $\delta=\delta_\textrm{eq}=1.5$ \citep{Paraschos24a} and estimated the emitting region size from our present data to be $\theta_{\textrm{mas}} \sim 0.15$\,mas \citep[including the correction factor discussed in][]{Marscher83}, based on the size reported in the works of \cite{Lister19}, \cite{Paraschos22}, and \cite{Kam24}, and then logarithmically interpolated to the turnover frequency \citep[see][for more details]{Pushkarev19}.
For the turnover parameters, works by \cite{Kim19} and \cite{Paraschos24a} place the turnover frequency and turnover flux in the ranges $\nu_m = 86-113\,\textrm{GHz}$ and $S^\prime_m = 5.6 - 9.0$\,Jy in the core region.
Lower values are expected downstream in a quiescent state, of the order of $\nu_m\sim20\,\textrm{GHz}$ \citep[e.g.][]{Benke26}.
In a flaring state, however, as is our case, the turnover parameters are expected to rise again, as shown in works by \cite{Valtaoja92}, \cite{Fromm16}, and \cite{Fuhrmann16}.
Furthermore, we computed the spectral indices $\alpha$ in the C3 area for the three epochs closest to the flare (i.e. the first three epochs of each frequency in Fig.~\ref{fig:BU}), following the procedure described in \cite{Paraschos21}.
The different frequency images used for the spectral analysis were not taken quasi-simultaneously, however, \C\ is known to not show variability of the order of a fortnight \citep[e.g.][]{Paraschos22, Park24}.
We found that before and after the flare $\alpha\sim-0.80$, while during the flare $\alpha\sim0.35$, hinting at higher turnover frequency values.
This spectral index trend is similar to the polarisation trend seen in Fig.~\ref{fig:SingleDish}.
We note that the uncertainties are of the order of $0.2$, which can influence the final turnover frequency values substantially.
Finally, we calculated the flux density at the turnover frequency $\nu_m=60\,\textrm{GHz}$ via the standard formula $S\propto\nu^\alpha$, using the measured 43\,GHz total intensity flux density of C3 during the flare (see Table~\ref{tab:Epochs}).
Under these conditions, $B_{\textrm{SSA}}\sim 3.0\,\textrm{G}$.

Then, continuing with the formalism presented in \cite{Marscher83}, the synchrotron self-Compton flux density can be approximated by the following equation:
\begin{align}
\begin{split}
        F_\textrm{SSC}(E_\textrm{keV}) = &\ d(\alpha) \ln{\left(\frac{\nu_2}{\nu_m}\right)}  \theta_\textrm{mas}^{-2(2\alpha+3)}  \nu_m^{-(3\alpha+5)} S_m^{\prime2(\alpha+2)} \\ 
        &\ E_\textrm{keV}^{-\alpha} \left(\frac{1+z}{\delta}\right)^{2(\alpha+2)}\ [\mu \textrm{Jy}]\label{eq:F}.
\end{split}
\end{align}
Here, $d(\alpha)$ is another dimensionless function, $\nu_2\equiv2.8\times10^6 B_\textrm{SSA}\gamma_2^2$ is the frequency cut-off to the synchrotron spectrum, with $\gamma^2_2=5\times10^5$ (we used a characteristic value from the range reported in \cite{Abdo09}), and $E_\textrm{keV}$ is the photon energy in the observer's frame (we assumed a typical value of $10^6$\,\textrm{keV}, for a TeV source such as \C).
Plugging in all the aforementioned values yields $F_\textrm{SSC} \sim 0.4\times10^{-5}\,\mu \textrm{Jy}$, which agrees well with the peak flare value measured by Fermi, since converting it to micro Jansky yields $F^{\textit{Fermi}}_{\textrm{SSC}} \sim 1.3\times10^{-5}\,\mu \textrm{Jy}$.
Finally, to get an estimate of the order of magnitude of the associated uncertainties and given the non-linearity of Eqs.~\ref{eq:bssa} and ~\ref{eq:F}, we implemented a Markov Chain Monte Carlo approach, for $S^\prime_m$, $\nu_m$, and $\theta_\textrm{mas}$.
Specifically, for $\nu_m$ we chose a rather agnostic approach of assuming a uniform distribution, between 20\,GHz and 60\,GHz, while for the other two parameters normal distributions (with an uncertainty of $\delta\theta_\textrm{mas}=[0.10, 0.4]\,\textrm{mas}$, based on geometrical model-fitting, see \citealt{Paraschos24c, Kam24}, and $\delta S^\prime_m = [1.1, 2.3]\,\textrm{Jy}$, based on the spectrum with the $\alpha$ used in the calculations above).
Our analysis resulted in [16th-84th] percentile limits of $\delta B_\textrm{SSA}=[0.02, 61.33]\,\textrm{G}$ and $F_\textrm{SSC} = [0, 380]\times10^{-5}\,\mu\textrm{Jy}$ respectively, emphasising the uncertainty of this calculation.
Expressed in terms of average and standard deviation, $B_\textrm{SSA}=10^{0.1\pm2.0}\,\textrm{G}$ and $F_\textrm{SSC}=10^{-5.4\pm3.5}\,\mu\textrm{Jy}$.

Second, we examined the behaviour of the VLBI EVPAs.
In our case, they remained at a consistent, perpendicular to the bulk jet flow orientation in the core region, while being variable close to the C3 region.
At the December 15 2024 epoch (see Fig.~\ref{fig:BU}), the EVPAs at the C3 region seem to follow the jet direction locally, similar to the X-ray results.
Their variability, however, in the adjacent epochs points towards turbulence, which could be the mechanism responsible for correlating the radio linear polarisation increase with the $\gamma$-ray emission.
Thus, taken our two-fold approach into account, in combination also with the insights of the accompanying X-ray analysis, we find that synchrotron self-Compton is indeed a viable emission mechanism.

\section{Conclusions} \label{sec:Conclusions}

In this work we presented VLBI observations of \C, taken during a $\gamma$-ray flare.
We performed a study to localise the origin of these $\gamma$-ray photons via polarimetric imaging of the VLBI epochs.
Our findings can be summarised as follows:
\begin{itemize}
    \item Analysing centimetre and millimetre VLBI epochs of \C\ before, during, and after a $\gamma$-ray flare revealed increased linearly polarised emission at $\sim1.5$\,parsec downstream of the central engine.
    \item Simultaneously taken X-ray measurements show a clear detection of polarised X-ray emission in the source, which, in combination with the radio data discussed here, provides a circumstantial hint towards the blazar zone being beyond the ultimate vicinity of the central engine and the up-scattering mechanism possibly being synchrotron self-Compton.
\end{itemize}

In summary, our results are in support of a far site scenario, in which $\gamma$-rays are produced beyond the broad line region of \C.
However, our work does not exclude the near site scenario also being at play during other periods of intense activity within the jet.

\begin{acknowledgements}
      We thank the anonymous referee for their valuable comments which greatly improved this manuscript.
      We thank Dan Homan for performing part of the MOJAVE data analysis used in this work.
      This research is supported by the European Research Council advanced grant “M2FINDERS - Mapping Magnetic Fields with INterferometry Down to Event hoRizon Scales” (Grant No. 101018682). 
      I. Liodakis was funded by the European Union ERC-2022-STG - BOOTES - 101076343. Views and opinions expressed are however those of the author(s) only and do not necessarily reflect those of the European Union or the European Research Council Executive Agency. Neither the European Union nor the granting authority can be held responsible for them.
      Y. Y. Kovalev was supported by the MuSES project, which has received funding from the European Union (ERC grant agreement No 101142396). Views and opinions expressed are however those of the author(s) only and do not necessarily reflect those of the European Union or ERCEA. Neither the European Union nor the granting authority can be held responsible for them.
      A. B. Pushkarev is supported in the framework of the State project `Science' by the Ministry of Science and Higher Education of the Russian Federation under the contract 075-15-2024-541.
      C. Casadio and D. \'{A}lvarez-Ortega acknowledge support from the European Research Council (ERC) under the Horizon ERC Grants 2021 programme under grant agreement No.101040021. 
      The University of New Hampshire group is supported in part by NASA Astrophysics Astrophysics Data Analysis Program grant 80NSSC24K0636. 
      The POLAMI observations reported here were carried out at the IRAM 30m Telescope. IRAM is supported by INSU/CNRS (France), MPG (Germany) and IGN (Spain).
      The Submillimeter Array (SMA) is a joint project between the Smithsonian Astrophysical Observatory and the Academia Sinica Institute of Astronomy and Astrophysics and is funded by the Smithsonian Institution and the Academia Sinica. Maunakea, the location of the SMA, is a culturally important site for the indigenous Hawaiian people; we are privileged to study the cosmos from its summit.
      The KVN is a facility operated by the Korea Astronomy and Space Science Institute. The KVN operations are supported by KREONET (Korea Research Environment Open NETwork) which is managed and operated by KISTI (Korea Institute of Science and Technology Information). S. Kang, S.-S. Lee, W. Y. Cheong, S.-H. Kim, and H.-W. Jeong were supported by the National Research Foundation of Korea (NRF) grant funded by the Korea government (MIST) (2020R1A2C2009003, RS-2025-00562700).
      The IAA-CSIC co-authors acknowledge financial support from the Spanish "Ministerio de Ciencia e Innovaci\'{o}n" (MCIN/AEI/ 10.13039/501100011033) through the Center of Excellence Severo Ochoa award for the Instituto de Astrof\'{i}isica de Andaluc\'{i}a-CSIC (CEX2021-001131-S), and through grants PID2019-107847RB-C44 and PID2022-139117NB-C44.
      This study makes use of VLBA data from the VLBA-BU Blazar Monitoring Program (BEAM-ME and VLBA-BU-BLAZAR;
      \url{http://www.bu.edu/blazars/BEAM-ME.html}), funded by NASA through the Fermi Guest Investigator Program. The VLBA is an instrument of the National Radio Astronomy Observatory. The National Radio Astronomy Observatory is a facility of the National Science Foundation operated by Associated Universities, Inc.
      This research has made use of data from the MOJAVE database that is maintained by the MOJAVE team \citep{Lister18}.
      This research has made use of the NASA/IPAC Extragalactic Database (NED), which is operated by the Jet Propulsion Laboratory, California Institute of Technology, under contract with the National Aeronautics and Space Administration. 
      This research has also made use of NASA's Astrophysics Data System Bibliographic Services. 
      Finally, this research made use of the following python packages: {\it numpy} \citep{Harris20}, {\it scipy} \citep{2020SciPy-NMeth}, {\it matplotlib} \citep{Hunter07}, {\it astropy} \citep{2013A&A...558A..33A, 2018AJ....156..123A} and {\it Uncertainties: a Python package for calculations with uncertainties.
      }
\end{acknowledgements}

\bibliographystyle{aa} 
\bibliography{aa59250-26} 

@ARTICLE{Backer87,
       author = {{Backer}, D.~C. and {Wright}, M.~C.~H. and {Plambeck}, R.~L. and {Carlstrom}, J.~E. and {Masson}, C.~R. and {Moffet}, A.~T. and {Readhead}, A.~C.~S. and {Woody}, D. and {Rogers}, A.~E.~E. and {Moran}, J.~M. and {Predmore}, C.~R. and {Dickman}, R.~L.},
        title = "{VLBI Structure of 3C 84 at 89 GHz}",
      journal = {\apj},
         year = 1987,
        month = nov,
       volume = {322},
        pages = {74},
          doi = {10.1086/165704},
       adsurl = {https://ui.adsabs.harvard.edu/abs/1987ApJ...322...74B}
}

@ARTICLE{Abdollahi23,
       author = {{Abdollahi}, S. and {Ajello}, M. and {Baldini}, L. and {Ballet}, J. and {Bastieri}, D. and {Becerra Gonzalez}, J. and {Bellazzini}, R. and {Berretta}, A. and {Bissaldi}, E. and {Bonino}, R. and {Brill}, A. and {Bruel}, P. and {Burns}, E. and {Buson}, S. and {Cameron}, R.~A. and {Caputo}, R. and {Caraveo}, P.~A. and {Cibrario}, N. and {Ciprini}, S. and {Cristarella Orestano}, P. and {Crnogorcevic}, M. and {Cutini}, S. and {D'Ammando}, F. and {De Gaetano}, S. and {Digel}, S.~W. and {Di Lalla}, N. and {Di Venere}, L. and {Dom{\'\i}nguez}, A. and {Ramazani}, V. Fallah and {Fegan}, S.~J. and {Ferrara}, E.~C. and {Fiori}, A. and {Fleischhack}, H. and {Franckowiak}, A. and {Fukazawa}, Y. and {Fusco}, P. and {Gammaldi}, V. and {Gargano}, F. and {Garrappa}, S. and {Gasbarra}, C. and {Gasparrini}, D. and {Giglietto}, N. and {Giordano}, F. and {Giroletti}, M. and {Green}, D. and {Grenier}, I.~A. and {Guiriec}, S. and {Gustafsson}, M. and {Hays}, E. and {Horan}, D. and {Hou}, X. and {J{\'o}hannesson}, G. and {Kerr}, M. and {Kocevski}, D. and {Kuss}, M. and {Latronico}, L. and {Li}, J. and {Liodakis}, I. and {Longo}, F. and {Loparco}, F. and {Lorusso}, L. and {Lott}, B. and {Lovellette}, M.~N. and {Lubrano}, P. and {Maldera}, S. and {Manfreda}, A. and {Mart{\'\i}-Devesa}, G. and {Mazziotta}, M.~N. and {Mereu}, I. and {Meyer}, M. and {Michelson}, P.~F. and {Mizuno}, T. and {Monzani}, M.~E. and {Morselli}, A. and {Moskalenko}, I.~V. and {Negro}, M. and {Omodei}, N. and {Orlando}, E. and {Ormes}, J.~F. and {Paneque}, D. and {Panzarini}, G. and {Perkins}, J.~S. and {Persic}, M. and {Pesce-Rollins}, M. and {Pillera}, R. and {Porter}, T.~A. and {Principe}, G. and {Racusin}, J.~L. and {Rain{\`o}}, S. and {Rando}, R. and {Rani}, B. and {Razzano}, M. and {Razzaque}, S. and {Reimer}, A. and {Reimer}, O. and {S{\'a}nchez-Conde}, M. and {Parkinson}, P.~M. Saz and {Scargle}, Jeff and {Scotton}, L. and {Serini}, D. and {Sgr{\`o}}, C. and {Siskind}, E.~J. and {Spandre}, G. and {Spinelli}, P. and {Suson}, D.~J. and {Tajima}, H. and {Thompson}, D.~J. and {Torres}, D.~F. and {Valverde}, J. and {Venters}, T. and {Wadiasingh}, Z. and {Wagner}, S. and {Wood}, K.},
        title = "{The Fermi-LAT Lightcurve Repository}",
      journal = {\apjs},
         year = 2023,
        month = apr,
       volume = {265},
       number = {2},
          eid = {31},
        pages = {31},
          doi = {10.3847/1538-4365/acbb6a},
archivePrefix = {arXiv},
       eprint = {2301.01607},
 primaryClass = {astro-ph.HE},
       adsurl = {https://ui.adsabs.harvard.edu/abs/2023ApJS..265...31A}
}

@ARTICLE{Paraschos25a,
       author = {{Paraschos}, G.~F.},
        title = "{A shocking outcome: Jet dynamics and polarimetric signatures of the multi-band flare in blazar OJ 248}",
      journal = {\aap},
         year = 2025,
        month = mar,
       volume = {695},
          eid = {L3},
        pages = {L3},
          doi = {10.1051/0004-6361/202553689},
archivePrefix = {arXiv},
       eprint = {2502.12232},
 primaryClass = {astro-ph.HE},
       adsurl = {https://ui.adsabs.harvard.edu/abs/2025A&A...695L...3P}
}

@ARTICLE{Paraschos25b,
       author = {{Paraschos}, G.~F. and {Mpisketzis}, V.},
        title = "{Unravelling the dynamics of cosmic vortices: Probing a Kelvin-Helmholtz instability in the jet of 3C 84}",
      journal = {\aap},
         year = 2025,
        month = apr,
       volume = {696},
          eid = {L7},
        pages = {L7},
          doi = {10.1051/0004-6361/202554201},
archivePrefix = {arXiv},
       eprint = {2503.16647},
 primaryClass = {astro-ph.HE},
       adsurl = {https://ui.adsabs.harvard.edu/abs/2025A&A...696L...7P}
}

@ARTICLE{Paraschos25c,
       author = {{Paraschos}, G.~F. and {Traianou}, E. and {Debbrecht}, L.~C. and {Liodakis}, I. and {Ros}, E.},
        title = "{Polarization as a Probe of Neutrino Emission from Blazars}",
      journal = {\apj},
         year = 2025,
        month = aug,
       volume = {989},
       number = {2},
          eid = {208},
        pages = {208},
          doi = {10.3847/1538-4357/adf110},
archivePrefix = {arXiv},
       eprint = {2507.16929},
 primaryClass = {astro-ph.HE},
       adsurl = {https://ui.adsabs.harvard.edu/abs/2025ApJ...989..208P}
}

@ARTICLE{Dotson15,
       author = {{Dotson}, Amanda and {Georganopoulos}, Markos and {Meyer}, Eileen T. and {McCann}, Kevin},
        title = "{On the Location of the 2009 GeV Flares of Blazar PKS 1510-089}",
      journal = {\apj},
         year = 2015,
        month = aug,
       volume = {809},
       number = {2},
          eid = {164},
        pages = {164},
          doi = {10.1088/0004-637X/809/2/164},
       adsurl = {https://ui.adsabs.harvard.edu/abs/2015ApJ...809..164D}
}

@ARTICLE{Lico17,
       author = {{Lico}, R. and {Giroletti}, M. and {Orienti}, M. and {Costamante}, L. and {Pavlidou}, V. and {D'Ammando}, F. and {Tavecchio}, F.},
        title = "{Exploring the connection between radio and GeV-TeV {\ensuremath{\gamma}}-ray emission in the 1FHL and 2FHL AGN samples}",
      journal = {\aap},
         year = 2017,
        month = oct,
       volume = {606},
          eid = {A138},
        pages = {A138},
          doi = {10.1051/0004-6361/201731116},
archivePrefix = {arXiv},
       eprint = {1708.06201},
 primaryClass = {astro-ph.HE},
       adsurl = {https://ui.adsabs.harvard.edu/abs/2017A&A...606A.138L}
}

@ARTICLE{Liodakis25,
       author = {{Liodakis}, Ioannis and {Chakraborty}, Sudip and {Marin}, Fr{\'e}d{\'e}ric and {Ehlert}, Steven R. and {Barnouin}, Thibault and {Kouch}, Pouya M. and {Nilsson}, Kari and {Lindfors}, Elina and {Pursimo}, Tapio and {Paraschos}, Georgios F. and {Middei}, Riccardo and {Trindade Falc{\~a}o}, Anna and {Jorstad}, Svetlana and {Agudo}, Iv{\'a}n and {Kovalev}, Yuri Y. and {Casey}, Jacob J. and {Di Gesu}, Laura and {Kaaret}, Philip and {Kim}, Dawoon E. and {Kislat}, Fabian and {Ratheesh}, Ajay and {Saade}, M. Lynne and {Tombesi}, Francesco and {Marscher}, Alan and {Jos{\'e} Aceituno}, Francisco and {Bonnoli}, Giacomo and {Casanova}, V{\'\i}ctor and {Emery}, Gabriel and {Escudero Pedrosa}, Juan and {Morcuende}, Daniel and {Otero-Santos}, Jorge and {Sota}, Alfredo and {Piirola}, Vilppu and {Bachev}, Rumen and {Strigachev}, Anton and {Borman}, George A. and {Grishina}, Tatiana S. and {Hagen-Thorn}, Vladimir A. and {Kopatskaya}, Evgenia N. and {Larionova}, Elena G. and {Morozova}, Daria A. and {Savchenko}, Sergey S. and {Shishkina}, Ekaterina V. and {Troitskiy}, Ivan S. and {Troitskaya}, Yulia V. and {Vasilyev}, Andrey A. and {Zhovtan}, Alexey V. and {Myserlis}, Ioannis and {Gurwell}, Mark and {Keating}, Garrett and {Rao}, Ramprasad and {Kang}, Sincheol and {Lee}, Sang-Sung and {Kim}, Sanghyun and {Yeon Cheong}, Whee and {Jeong}, Hyeon-Woo and {Song}, Chanwoo and {Li}, Shan and {Nam}, Myeong-Seok and {{\'A}lvarez-Ortega}, Diego and {Casadio}, Carolina and {Angelakis}, Emmanouil and {Kraus}, Alexander and {Jormanainen}, Jenni and {Fallah Ramazani}, Vandad and {Chen}, Chien-Ting and {Costa}, Enrico and {Churazov}, Eugene and {Ferrazzoli}, Riccardo and {Galanti}, Giorgio and {Khabibullin}, Ildar and {O'Dell}, Stephen L. and {Pacciani}, Luigi and {Roncadelli}, Marco and {Roberts}, Oliver J. and {Soffitta}, Paolo and {Swartz}, Douglas A. and {Tavecchio}, Fabrizio and {Weisskopf}, Martin C. and {Zhuravleva}, Irina},
        title = "{Detection of Compton Scattering in the Jet of 3C 84}",
      journal = {\apjl},
         year = 2025,
        month = nov,
       volume = {994},
       number = {1},
          eid = {L9},
        pages = {L9},
          doi = {10.3847/2041-8213/ae157d},
       adsurl = {https://ui.adsabs.harvard.edu/abs/2025ApJ...994L...9L}
}

@ARTICLE{Biteau12,
       author = {{Biteau}, J. and {Giebels}, B.},
        title = "{The minijets-in-a-jet statistical model and the rms-flux correlation}",
      journal = {\aap},
         year = 2012,
        month = dec,
       volume = {548},
          eid = {A123},
        pages = {A123},
          doi = {10.1051/0004-6361/201220056},
archivePrefix = {arXiv},
       eprint = {1210.2045},
 primaryClass = {astro-ph.HE},
       adsurl = {https://ui.adsabs.harvard.edu/abs/2012A&A...548A.123B}
}

@ARTICLE{Weisskopf2022,
       author = {{Weisskopf}, Martin C. and {Soffitta}, Paolo and {Baldini}, Luca and {Ramsey}, Brian D. and {O'Dell}, Stephen L. and {Romani}, Roger W. and {Matt}, Giorgio and {Deininger}, William D. and {Baumgartner}, Wayne H. and {Bellazzini}, Ronaldo and {Costa}, Enrico and {Kolodziejczak}, Jeffery J. and {Latronico}, Luca and {Marshall}, Herman L. and {Muleri}, Fabio and {Bongiorno}, Stephen D. and {Tennant}, Allyn and {Bucciantini}, Niccolo and {Dovciak}, Michal and {Marin}, Frederic and {Marscher}, Alan and {Poutanen}, Juri and {Slane}, Pat and {Turolla}, Roberto and {Kalinowski}, William and {Di Marco}, Alessandro and {Fabiani}, Sergio and {Minuti}, Massimo and {La Monaca}, Fabio and {Pinchera}, Michele and {Rankin}, John and {Sgro'}, Carmelo and {Trois}, Alessio and {Xie}, Fei and {Alexander}, Cheryl and {Allen}, D. Zachery and {Amici}, Fabrizio and {Andersen}, Jason and {Antonelli}, Angelo and {Antoniak}, Spencer and {Attin{\`a}}, Primo and {Barbanera}, Mattia and {Bachetti}, Matteo and {Baggett}, Randy M. and {Bladt}, Jeff and {Brez}, Alessandro and {Bonino}, Raffaella and {Boree}, Christopher and {Borotto}, Fabio and {Breeding}, Shawn and {Brienza}, Daniele and {Bygott}, H. Kyle and {Caporale}, Ciro and {Cardelli}, Claudia and {Carpentiero}, Rita and {Castellano}, Simone and {Castronuovo}, Marco and {Cavalli}, Luca and {Cavazzuti}, Elisabetta and {Ceccanti}, Marco and {Centrone}, Mauro and {Citraro}, Saverio and {D'Amico}, Fabio and {D'Alba}, Elisa and {Di Gesu}, Laura and {Del Monte}, Ettore and {Dietz}, Kurtis L. and {Di Lalla}, Niccolo' and {Persio}, Giuseppe Di and {Dolan}, David and {Donnarumma}, Immacolata and {Evangelista}, Yuri and {Ferrant}, Kevin and {Ferrazzoli}, Riccardo and {Ferrie}, MacKenzie and {Footdale}, Joseph and {Forsyth}, Brent and {Foster}, Michelle and {Garelick}, Benjamin and {Gunji}, Shuichi and {Gurnee}, Eli and {Head}, Michael and {Hibbard}, Grant and {Johnson}, Samantha and {Kelly}, Erik and {Kilaru}, Kiranmayee and {Lefevre}, Carlo and {Roy}, Shelley Le and {Loffredo}, Pasqualino and {Lorenzi}, Paolo and {Lucchesi}, Leonardo and {Maddox}, Tyler and {Magazzu}, Guido and {Maldera}, Simone and {Manfreda}, Alberto and {Mangraviti}, Elio and {Marengo}, Marco and {Marrocchesi}, Alessandra and {Massaro}, Francesco and {Mauger}, David and {McCracken}, Jeffrey and {McEachen}, Michael and {Mize}, Rondal and {Mereu}, Paolo and {Mitchell}, Scott and {Mitsuishi}, Ikuyuki and {Morbidini}, Alfredo and {Mosti}, Federico and {Nasimi}, Hikmat and {Negri}, Barbara and {Negro}, Michela and {Nguyen}, Toan and {Nitschke}, Isaac and {Nuti}, Alessio and {Onizuka}, Mitch and {Oppedisano}, Chiara and {Orsini}, Leonardo and {Osborne}, Darren and {Pacheco}, Richard and {Paggi}, Alessandro and {Painter}, Will and {Pavelitz}, Steven D. and {Pentz}, Christina and {Piazzolla}, Raffaele and {Perri}, Matteo and {Pesce-Rollins}, Melissa and {Peterson}, Colin and {Pilia}, Maura and {Profeti}, Alessandro and {Puccetti}, Simonetta and {Ranganathan}, Jaganathan and {Ratheesh}, Ajay and {Reedy}, Lee and {Root}, Noah and {Rubini}, Alda and {Ruswick}, Stephanie and {Sanchez}, Javier and {Sarra}, Paolo and {Santoli}, Francesco and {Scalise}, Emanuele and {Sciortino}, Andrea and {Schroeder}, Christopher and {Seek}, Tim and {Sosdian}, Kalie and {Spandre}, Gloria and {Speegle}, Chet O. and {Tamagawa}, Toru and {Tardiola}, Marcello and {Tobia}, Antonino and {Thomas}, Nicholas E. and {Valerie}, Robert and {Vimercati}, Marco and {Walden}, Amy L. and {Weddendorf}, Bruce and {Wedmore}, Jeffrey and {Welch}, David and {Zanetti}, Davide and {Zanetti}, Francesco},
        title = "{The Imaging X-Ray Polarimetry Explorer (IXPE): Pre-Launch}",
      journal = {Journal of Astronomical Telescopes, Instruments, and Systems},
         year = 2022,
        month = apr,
       volume = {8},
       number = {2},
          eid = {026002},
        pages = {026002},
          doi = {10.1117/1.JATIS.8.2.026002},
archivePrefix = {arXiv},
       eprint = {2112.01269},
 primaryClass = {astro-ph.IM},
       adsurl = {https://ui.adsabs.harvard.edu/abs/2022JATIS...8b6002W}
}

@INPROCEEDINGS{Soffitta23,
       author = {{Soffitta}, Paolo and {Baldini}, Luca and {Baumgartner}, Wayne and {Bellazzini}, Ronaldo and {Bongiorno}, Stephen D. and {Bucciantini}, Niccol{\`o} and {Costa}, Enrico and {Dov{\v{c}}iak}, Michal and {Ehlert}, Steven and {Kaaret}, Philip E. and {Kolodziejczak}, Jeffery J. and {Latronico}, Luca and {Marin}, Fr{\'e}d{\'e}ric and {Marscher}, Alan P. and {Marshall}, Herman L. and {Matt}, Giorgio and {Muleri}, Fabio and {O'Dell}, Stephen L. and {Poutanen}, Juri and {Ramsey}, Brian and {Romani}, Roger W. and {Slane}, Patrick and {Tennant}, Allyn F. and {Turolla}, Roberto and {Weisskopf}, Martin C. and {Agudo}, Iv{\'a}n. and {Antonelli}, Lucio Angelo and {Bachetti}, Matteo and {Bianchi}, Stefano and {Bonino}, Raffaella and {Brez}, Alessandro and {Capitanio}, Fiamma and {Castellano}, Simone and {Cavazzuti}, Elisabetta and {Chen}, Chiel-Ting and {Ciprini}, Stefano and {De Rosa}, Alessandra and {Del Monte}, Ettore and {Di Gesu}, Laura and {Di Lalla}, Niccol{\`o} and {Di Marco}, Alessandro and {Donnarumma}, Immacolata and {Doroshenko}, Victor and {Enoto}, Teruaki and {Evangelista}, Yuri and {Fabiani}, Sergio and {Ferrazzoli}, Riccardo and {Garcia}, Javier A. and {Gunji}, Shuichi and {Hayashida}, Kiyoshi and {Heyl}, Jeremy and {Iwakiri}, Wataru and {Jorstad}, Svetlana G. and {Karas}, Vladimir and {Kislat}, Fabian and {Kitaguchi}, Takao and {Krawczynski}, Henric and {La Monaca}, Fabio and {Liodakis}, Ioannis and {Maldera}, Simone and {Manfreda}, Alberto and {Marinucci}, Andrea and {Massaro}, Francesco and {Mitsuishi}, Ikuyuki and {Mizuno}, Tsunefumi and {Negro}, Michela and {Ng}, C. -Y. and {Omodei}, Nicola and {Oppedisano}, Chiara and {Papitto}, Alessandro and {Pavlov}, George G. and {Peirson}, Abel Lawrence and {Perri}, Matteo and {Pesce-Rollins}, Melissa and {Petrucci}, Pierre-Olivier and {Pilia}, Maura and {Possenti}, Andrea and {Puccetti}, Simonetta and {Rankin}, John and {Ratheesh}, Ajay and {Roberts}, Oliver J. and {Sgr{\`o}}, Carmelo and {Spandre}, Gloria and {Swartz}, Douglas A. and {Tamagawa}, Toru and {Tavecchio}, Fabrizio and {Taverna}, Roberto and {Tawara}, Yuzuru and {Thomas}, Nicholas E. and {Tombesi}, Francesco and {Trois}, Alessio and {Tsygankov}, Sergey S. and {Vink}, Jacco and {Wu}, Kinwah and {Xie}, Fei and {Zane}, Silvia},
        title = "{The Imaging x-ray polarimetry explorer (IXPE) at last!}",
    booktitle = {UV, X-Ray, and Gamma-Ray Space Instrumentation for Astronomy XXIII},
         year = 2023,
       editor = {{Siegmund}, Oswald H. and {Hoadley}, Keri},
       series = {Society of Photo-Optical Instrumentation Engineers (SPIE) Conference Series},
       volume = {12678},
        month = oct,
          eid = {1267803},
        pages = {1267803},
          doi = {10.1117/12.2677296},
       adsurl = {https://ui.adsabs.harvard.edu/abs/2023SPIE12678E..03S}
}

@ARTICLE{Abdo09,
       author = {{Abdo}, A.~A. and {Ackermann}, M. and {Ajello}, M. and {Asano}, K. and {Baldini}, L. and {Ballet}, J. and {Barbiellini}, G. and {Bastieri}, D. and {Baughman}, B.~M. and {Bechtol}, K. and {Bellazzini}, R. and {Blandford}, R.~D. and {Bloom}, E.~D. and {Bonamente}, E. and {Borgland}, A.~W. and {Bregeon}, J. and {Brez}, A. and {Brigida}, M. and {Bruel}, P. and {Burnett}, T.~H. and {Caliandro}, G.~A. and {Cameron}, R.~A. and {Caraveo}, P.~A. and {Casandjian}, J.~M. and {Cavazzuti}, E. and {Cecchi}, C. and {Celotti}, A. and {Chekhtman}, A. and {Cheung}, C.~C. and {Chiang}, J. and {Ciprini}, S. and {Claus}, R. and {Cohen-Tanugi}, J. and {Colafrancesco}, S. and {Cominsky}, L.~R. and {Conrad}, J. and {Costamante}, L. and {Dermer}, C.~D. and {de Angelis}, A. and {de Palma}, F. and {Digel}, S.~W. and {Donato}, D. and {do Couto e Silva}, E. and {Drell}, P.~S. and {Dubois}, R. and {Dumora}, D. and {Farnier}, C. and {Favuzzi}, C. and {Finke}, J. and {Focke}, W.~B. and {Frailis}, M. and {Fukazawa}, Y. and {Funk}, S. and {Fusco}, P. and {Gargano}, F. and {Georganopoulos}, M. and {Germani}, S. and {Giebels}, B. and {Giglietto}, N. and {Giordano}, F. and {Glanzman}, T. and {Grenier}, I.~A. and {Grondin}, M. -H. and {Grove}, J.~E. and {Guillemot}, L. and {Guiriec}, S. and {Hanabata}, Y. and {Harding}, A.~K. and {Hartman}, R.~C. and {Hayashida}, M. and {Hays}, E. and {Hughes}, R.~E. and {J{\'o}hannesson}, G. and {Johnson}, A.~S. and {Johnson}, R.~P. and {Johnson}, W.~N. and {Kadler}, M. and {Kamae}, T. and {Kanai}, Y. and {Katagiri}, H. and {Kataoka}, J. and {Kawai}, N. and {Kerr}, M. and {Kn{\"o}dlseder}, J. and {Kuehn}, F. and {Kuss}, M. and {Latronico}, L. and {Lemoine-Goumard}, M. and {Longo}, F. and {Loparco}, F. and {Lott}, B. and {Lovellette}, M.~N. and {Lubrano}, P. and {Madejski}, G.~M. and {Makeev}, A. and {Mazziotta}, M.~N. and {McEnery}, J.~E. and {Meurer}, C. and {Michelson}, P.~F. and {Mitthumsiri}, W. and {Mizuno}, T. and {Moiseev}, A.~A. and {Monte}, C. and {Monzani}, M.~E. and {Morselli}, A. and {Moskalenko}, I.~V. and {Murgia}, S. and {Nakamori}, T. and {Nolan}, P.~L. and {Norris}, J.~P. and {Nuss}, E. and {Ohsugi}, T. and {Omodei}, N. and {Orlando}, E. and {Ormes}, J.~F. and {Paneque}, D. and {Panetta}, J.~H. and {Parent}, D. and {Pepe}, M. and {Pesce-Rollins}, M. and {Piron}, F. and {Porter}, T.~A. and {Rain{\`o}}, S. and {Razzano}, M. and {Reimer}, A. and {Reimer}, O. and {Reposeur}, T. and {Ritz}, S. and {Rodriguez}, A.~Y. and {Romani}, R.~W. and {Ryde}, F. and {Sadrozinski}, H.~F. -W. and {Sambruna}, R. and {Sanchez}, D. and {Sander}, A. and {Sato}, R. and {Parkinson}, P.~M. Saz and {Sgr{\`o}}, C. and {Smith}, D.~A. and {Smith}, P.~D. and {Spandre}, G. and {Spinelli}, P. and {Starck}, J. -L. and {Strickman}, M.~S. and {Strong}, A.~W. and {Suson}, D.~J. and {Tajima}, H. and {Takahashi}, H. and {Takahashi}, T. and {Tanaka}, T. and {Taylor}, G.~B. and {Thayer}, J.~G. and {Thompson}, D.~J. and {Torres}, D.~F. and {Tosti}, G. and {Uchiyama}, Y. and {Usher}, T.~L. and {Vilchez}, N. and {Vitale}, V. and {Waite}, A.~P. and {Wood}, K.~S. and {Ylinen}, T. and {Ziegler}, M. and {Aller}, H.~D. and {Aller}, M.~F. and {Kellermann}, K.~I. and {Kovalev}, Y.~Y. and {Kovalev}, Yu. A. and {Lister}, M.~L. and {Pushkarev}, A.~B.},
        title = "{Fermi Discovery of Gamma-ray Emission from NGC 1275}",
      journal = {\apj},
         year = 2009,
        month = jul,
       volume = {699},
       number = {1},
        pages = {31-39},
          doi = {10.1088/0004-637X/699/1/31},
archivePrefix = {arXiv},
       eprint = {0904.1904},
 primaryClass = {astro-ph.HE},
       adsurl = {https://ui.adsabs.harvard.edu/abs/2009ApJ...699...31A}
}

@ARTICLE{Agudo11,
       author = {{Agudo}, Iv{\'a}n and {Jorstad}, Svetlana G. and {Marscher}, Alan P. and {Larionov}, Valeri M. and {G{\'o}mez}, Jos{\'e} L. and {L{\"a}hteenm{\"a}ki}, Anne and {Gurwell}, Mark and {Smith}, Paul S. and {Wiesemeyer}, Helmut and {Thum}, Clemens and {Heidt}, Jochen and {Blinov}, Dmitriy A. and {D'Arcangelo}, Francesca D. and {Hagen-Thorn}, Vladimir A. and {Morozova}, Daria A. and {Nieppola}, Elina and {Roca-Sogorb}, Mar and {Schmidt}, Gary D. and {Taylor}, Brian and {Tornikoski}, Merja and {Troitsky}, Ivan S.},
        title = "{Location of {\ensuremath{\gamma}}-ray Flare Emission in the Jet of the BL Lacertae Object OJ287 More than 14 pc from the Central Engine}",
      journal = {\apjl},
         year = 2011,
        month = jan,
       volume = {726},
       number = {1},
          eid = {L13},
        pages = {L13},
          doi = {10.1088/2041-8205/726/1/L13},
archivePrefix = {arXiv},
       eprint = {1011.6454},
 primaryClass = {astro-ph.CO},
       adsurl = {https://ui.adsabs.harvard.edu/abs/2011ApJ...726L..13A}
}

@ARTICLE{Agudo18a,
       author = {{Agudo}, Iv{\'a}n and {Thum}, Clemens and {Molina}, Sol N. and {Casadio}, Carolina and {Wiesemeyer}, Helmut and {Morris}, David and {Paubert}, Gabriel and {G{\'o}mez}, Jos{\'e} L. and {Kramer}, Carsten},
        title = "{POLAMI: Polarimetric Monitoring of AGN at Millimetre Wavelengths - I. The programme, calibration and calibrator data products}",
      journal = {\mnras},
         year = 2018,
        month = feb,
       volume = {474},
       number = {2},
        pages = {1427-1435},
          doi = {10.1093/mnras/stx2435},
archivePrefix = {arXiv},
       eprint = {1709.08742},
 primaryClass = {astro-ph.GA},
       adsurl = {https://ui.adsabs.harvard.edu/abs/2018MNRAS.474.1427A}
}

@ARTICLE{2013A&A...558A..33A,
       author = {{Astropy Collaboration} and {Robitaille}, Thomas P. and {Tollerud}, Erik J. and {Greenfield}, Perry and {Droettboom}, Michael and {Bray}, Erik and {Aldcroft}, Tom and {Davis}, Matt and {Ginsburg}, Adam and {Price-Whelan}, Adrian M. and {Kerzendorf}, Wolfgang E. and {Conley}, Alexander and {Crighton}, Neil and {Barbary}, Kyle and {Muna}, Demitri and {Ferguson}, Henry and {Grollier}, Fr{\'e}d{\'e}ric and {Parikh}, Madhura M. and {Nair}, Prasanth H. and {Unther}, Hans M. and {Deil}, Christoph and {Woillez}, Julien and {Conseil}, Simon and {Kramer}, Roban and {Turner}, James E.~H. and {Singer}, Leo and {Fox}, Ryan and {Weaver}, Benjamin A. and {Zabalza}, Victor and {Edwards}, Zachary I. and {Azalee Bostroem}, K. and {Burke}, D.~J. and {Casey}, Andrew R. and {Crawford}, Steven M. and {Dencheva}, Nadia and {Ely}, Justin and {Jenness}, Tim and {Labrie}, Kathleen and {Lim}, Pey Lian and {Pierfederici}, Francesco and {Pontzen}, Andrew and {Ptak}, Andy and {Refsdal}, Brian and {Servillat}, Mathieu and {Streicher}, Ole},
        title = "{Astropy: A community Python package for astronomy}",
      journal = {\aap},
         year = 2013,
        month = oct,
       volume = {558},
          eid = {A33},
        pages = {A33},
          doi = {10.1051/0004-6361/201322068},
archivePrefix = {arXiv},
       eprint = {1307.6212},
 primaryClass = {astro-ph.IM},
       adsurl = {https://ui.adsabs.harvard.edu/abs/2013A&A...558A..33A}
}

@ARTICLE{2018AJ....156..123A,
       author = {{Astropy Collaboration} and {Price-Whelan}, A.~M. and {Sip{\H{o}}cz}, B.~M. and {G{\"u}nther}, H.~M. and {Lim}, P.~L. and {Crawford}, S.~M. and {Conseil}, S. and {Shupe}, D.~L. and {Craig}, M.~W. and {Dencheva}, N. and {Ginsburg}, A. and {VanderPlas}, J.~T. and {Bradley}, L.~D. and {P{\'e}rez-Su{\'a}rez}, D. and {de Val-Borro}, M. and {Aldcroft}, T.~L. and {Cruz}, K.~L. and {Robitaille}, T.~P. and {Tollerud}, E.~J. and {Ardelean}, C. and {Babej}, T. and {Bach}, Y.~P. and {Bachetti}, M. and {Bakanov}, A.~V. and {Bamford}, S.~P. and {Barentsen}, G. and {Barmby}, P. and {Baumbach}, A. and {Berry}, K.~L. and {Biscani}, F. and {Boquien}, M. and {Bostroem}, K.~A. and {Bouma}, L.~G. and {Brammer}, G.~B. and {Bray}, E.~M. and {Breytenbach}, H. and {Buddelmeijer}, H. and {Burke}, D.~J. and {Calderone}, G. and {Cano Rodr{\'\i}guez}, J.~L. and {Cara}, M. and {Cardoso}, J.~V.~M. and {Cheedella}, S. and {Copin}, Y. and {Corrales}, L. and {Crichton}, D. and {D'Avella}, D. and {Deil}, C. and {Depagne}, {\'E}. and {Dietrich}, J.~P. and {Donath}, A. and {Droettboom}, M. and {Earl}, N. and {Erben}, T. and {Fabbro}, S. and {Ferreira}, L.~A. and {Finethy}, T. and {Fox}, R.~T. and {Garrison}, L.~H. and {Gibbons}, S.~L.~J. and {Goldstein}, D.~A. and {Gommers}, R. and {Greco}, J.~P. and {Greenfield}, P. and {Groener}, A.~M. and {Grollier}, F. and {Hagen}, A. and {Hirst}, P. and {Homeier}, D. and {Horton}, A.~J. and {Hosseinzadeh}, G. and {Hu}, L. and {Hunkeler}, J.~S. and {Ivezi{\'c}}, {\v{Z}}. and {Jain}, A. and {Jenness}, T. and {Kanarek}, G. and {Kendrew}, S. and {Kern}, N.~S. and {Kerzendorf}, W.~E. and {Khvalko}, A. and {King}, J. and {Kirkby}, D. and {Kulkarni}, A.~M. and {Kumar}, A. and {Lee}, A. and {Lenz}, D. and {Littlefair}, S.~P. and {Ma}, Z. and {Macleod}, D.~M. and {Mastropietro}, M. and {McCully}, C. and {Montagnac}, S. and {Morris}, B.~M. and {Mueller}, M. and {Mumford}, S.~J. and {Muna}, D. and {Murphy}, N.~A. and {Nelson}, S. and {Nguyen}, G.~H. and {Ninan}, J.~P. and {N{\"o}the}, M. and {Ogaz}, S. and {Oh}, S. and {Parejko}, J.~K. and {Parley}, N. and {Pascual}, S. and {Patil}, R. and {Patil}, A.~A. and {Plunkett}, A.~L. and {Prochaska}, J.~X. and {Rastogi}, T. and {Reddy Janga}, V. and {Sabater}, J. and {Sakurikar}, P. and {Seifert}, M. and {Sherbert}, L.~E. and {Sherwood-Taylor}, H. and {Shih}, A.~Y. and {Sick}, J. and {Silbiger}, M.~T. and {Singanamalla}, S. and {Singer}, L.~P. and {Sladen}, P.~H. and {Sooley}, K.~A. and {Sornarajah}, S. and {Streicher}, O. and {Teuben}, P. and {Thomas}, S.~W. and {Tremblay}, G.~R. and {Turner}, J.~E.~H. and {Terr{\'o}n}, V. and {van Kerkwijk}, M.~H. and {de la Vega}, A. and {Watkins}, L.~L. and {Weaver}, B.~A. and {Whitmore}, J.~B. and {Woillez}, J. and {Zabalza}, V. and {Astropy Contributors}},
        title = "{The Astropy Project: Building an Open-science Project and Status of the v2.0 Core Package}",
      journal = {\aj},
         year = 2018,
        month = sep,
       volume = {156},
       number = {3},
          eid = {123},
        pages = {123},
          doi = {10.3847/1538-3881/aabc4f},
archivePrefix = {arXiv},
       eprint = {1801.02634},
 primaryClass = {astro-ph.IM},
       adsurl = {https://ui.adsabs.harvard.edu/abs/2018AJ....156..123A}
}

@ARTICLE{Blandford95,
       author = {{Blandford}, R.~D. and {Levinson}, A.},
        title = "{Pair Cascades in Extragalactic Jets. I. Gamma Rays}",
      journal = {\apj},
         year = 1995,
        month = mar,
       volume = {441},
        pages = {79},
          doi = {10.1086/175338},
       adsurl = {https://ui.adsabs.harvard.edu/abs/1995ApJ...441...79B}
}

@ARTICLE{Chael16,
       author = {{Chael}, Andrew A. and {Johnson}, Michael D. and {Narayan}, Ramesh and {Doeleman}, Sheperd S. and {Wardle}, John F.~C. and {Bouman}, Katherine L.},
        title = "{High-resolution Linear Polarimetric Imaging for the Event Horizon Telescope}",
      journal = {\apj},
         year = 2016,
        month = sep,
       volume = {829},
       number = {1},
          eid = {11},
        pages = {11},
          doi = {10.3847/0004-637X/829/1/11},
archivePrefix = {arXiv},
       eprint = {1605.06156},
 primaryClass = {astro-ph.IM},
       adsurl = {https://ui.adsabs.harvard.edu/abs/2016ApJ...829...11C}
}

@ARTICLE{Chael18,
       author = {{Chael}, Andrew A. and {Johnson}, Michael D. and {Bouman}, Katherine L. and {Blackburn}, Lindy L. and {Akiyama}, Kazunori and {Narayan}, Ramesh},
        title = "{Interferometric Imaging Directly with Closure Phases and Closure Amplitudes}",
      journal = {\apj},
         year = 2018,
        month = apr,
       volume = {857},
       number = {1},
          eid = {23},
        pages = {23},
          doi = {10.3847/1538-4357/aab6a8},
archivePrefix = {arXiv},
       eprint = {1803.07088},
 primaryClass = {astro-ph.IM},
       adsurl = {https://ui.adsabs.harvard.edu/abs/2018ApJ...857...23C}
}

@ARTICLE{EHT19a,
       author = {{EHTC} and {Akiyama}, Kazunori and {Alberdi}, Antxon and {Alef}, Walter and {Asada}, Keiichi and {Azulay}, Rebecca and {Baczko}, Anne-Kathrin and {Ball}, David and {Balokovi{\'c}}, Mislav and {Barrett}, John and {Bintley}, Dan and {Blackburn}, Lindy and {Boland}, Wilfred and {Bouman}, Katherine L. and {Bower}, Geoffrey C. and {Bremer}, Michael and {Brinkerink}, Christiaan D. and {Brissenden}, Roger and {Britzen}, Silke and {Broderick}, Avery E. and {Broguiere}, Dominique and {Bronzwaer}, Thomas and {Byun}, Do-Young and {Carlstrom}, John E. and {Chael}, Andrew and {Chan}, Chi-kwan and {Chatterjee}, Shami and {Chatterjee}, Koushik and {Chen}, Ming-Tang and {Chen}, Yongjun and {Cho}, Ilje and {Christian}, Pierre and {Conway}, John E. and {Cordes}, James M. and {Crew}, Geoffrey B. and {Cui}, Yuzhu and {Davelaar}, Jordy and {De Laurentis}, Mariafelicia and {Deane}, Roger and {Dempsey}, Jessica and {Desvignes}, Gregory and {Dexter}, Jason and {Doeleman}, Sheperd S. and {Eatough}, Ralph P. and {Falcke}, Heino and {Fish}, Vincent L. and {Fomalont}, Ed and {Fraga-Encinas}, Raquel and {Freeman}, William T. and {Friberg}, Per and {Fromm}, Christian M. and {G{\'o}mez}, Jos{\'e} L. and {Galison}, Peter and {Gammie}, Charles F. and {Garc{\'\i}a}, Roberto and {Gentaz}, Olivier and {Georgiev}, Boris and {Goddi}, Ciriaco and {Gold}, Roman and {Gu}, Minfeng and {Gurwell}, Mark and {Hada}, Kazuhiro and {Hecht}, Michael H. and {Hesper}, Ronald and {Ho}, Luis C. and {Ho}, Paul and {Honma}, Mareki and {Huang}, Chih-Wei L. and {Huang}, Lei and {Hughes}, David H. and {Ikeda}, Shiro and {Inoue}, Makoto and {Issaoun}, Sara and {James}, David J. and {Jannuzi}, Buell T. and {Janssen}, Michael and {Jeter}, Britton and {Jiang}, Wu and {Johnson}, Michael D. and {Jorstad}, Svetlana and {Jung}, Taehyun and {Karami}, Mansour and {Karuppusamy}, Ramesh and {Kawashima}, Tomohisa and {Keating}, Garrett K. and {Kettenis}, Mark and {Kim}, Jae-Young and {Kim}, Junhan and {Kim}, Jongsoo and {Kino}, Motoki and {Koay}, Jun Yi and {Koch}, Patrick M. and {Koyama}, Shoko and {Kramer}, Michael and {Kramer}, Carsten and {Krichbaum}, Thomas P. and {Kuo}, Cheng-Yu and {Lauer}, Tod R. and {Lee}, Sang-Sung and {Li}, Yan-Rong and {Li}, Zhiyuan and {Lindqvist}, Michael and {Liu}, Kuo and {Liuzzo}, Elisabetta and {Lo}, Wen-Ping and {Lobanov}, Andrei P. and {Loinard}, Laurent and {Lonsdale}, Colin and {Lu}, Ru-Sen and {MacDonald}, Nicholas R. and {Mao}, Jirong and {Markoff}, Sera and {Marrone}, Daniel P. and {Marscher}, Alan P. and {Mart{\'\i}-Vidal}, Iv{\'a}n and {Matsushita}, Satoki and {Matthews}, Lynn D. and {Medeiros}, Lia and {Menten}, Karl M. and {Mizuno}, Yosuke and {Mizuno}, Izumi and {Moran}, James M. and {Moriyama}, Kotaro and {Moscibrodzka}, Monika and {M{\"u}ller}, Cornelia and {Nagai}, Hiroshi and {Nagar}, Neil M. and {Nakamura}, Masanori and {Narayan}, Ramesh and {Narayanan}, Gopal and {Natarajan}, Iniyan and {Neri}, Roberto and {Ni}, Chunchong and {Noutsos}, Aristeidis and {Okino}, Hiroki and {Olivares}, H{\'e}ctor and {Ortiz-Le{\'o}n}, Gisela N. and {Oyama}, Tomoaki and {{\"O}zel}, Feryal and {Palumbo}, Daniel C.~M. and {Patel}, Nimesh and {Pen}, Ue-Li and {Pesce}, Dominic W. and {Pi{\'e}tu}, Vincent and {Plambeck}, Richard and {PopStefanija}, Aleksandar and {Porth}, Oliver and {Prather}, Ben and {Preciado-L{\'o}pez}, Jorge A. and {Psaltis}, Dimitrios and {Pu}, Hung-Yi and {Ramakrishnan}, Venkatessh and {Rao}, Ramprasad and {Rawlings}, Mark G. and {Raymond}, Alexander W. and {Rezzolla}, Luciano and {Ripperda}, Bart and {Roelofs}, Freek and {Rogers}, Alan and {Ros}, Eduardo and {Rose}, Mel and {Roshanineshat}, Arash and {Rottmann}, Helge and {Roy}, Alan L. and {Ruszczyk}, Chet and {Ryan}, Benjamin R. and {Rygl}, Kazi L.~J. and {S{\'a}nchez}, Salvador and {S{\'a}nchez-Arguelles}, David and {Sasada}, Mahito and {Savolainen}, Tuomas and {Schloerb}, F. Peter and {Schuster}, Karl-Friedrich and {Shao}, Lijing and {Shen}, Zhiqiang and {Small}, Des and {Sohn}, Bong Won and {SooHoo}, Jason and {Tazaki}, Fumie and {Tiede}, Paul and {Tilanus}, Remo P.~J. and {Titus}, Michael and {Toma}, Kenji and {Torne}, Pablo and {Trent}, Tyler and {Trippe}, Sascha and {Tsuda}, Shuichiro and {van Bemmel}, Ilse and {van Langevelde}, Huib Jan and {van Rossum}, Daniel R. and {Wagner}, Jan and {Wardle}, John and {Weintroub}, Jonathan and {Wex}, Norbert and {Wharton}, Robert and {Wielgus}, Maciek and {Wong}, George N. and {Wu}, Qingwen and {Young}, Ken and {Young}, Andr{\'e} and {Younsi}, Ziri and {Yuan}, Feng and {Yuan}, Ye-Fei and {Zensus}, J. Anton and {Zhao}, Guangyao and {Zhao}, Shan-Shan and {Zhu}, Ziyan and {Algaba}, Juan-Carlos and {Allardi}, Alexander and {Amestica}, Rodrigo and {Anczarski}, Jadyn and {Bach}, Uwe and {Baganoff}, Frederick K. and {Beaudoin}, Christopher and {Benson}, Bradford A. and {Berthold}, Ryan and {Blanchard}, Jay M. and {Blundell}, Ray and {Bustamente}, Sandra and {Cappallo}, Roger and {Castillo-Dom{\'\i}nguez}, Edgar and {Chang}, Chih-Cheng and {Chang}, Shu-Hao and {Chang}, Song-Chu and {Chen}, Chung-Chen and {Chilson}, Ryan and {Chuter}, Tim C. and {C{\'o}rdova Rosado}, Rodrigo and {Coulson}, Iain M. and {Crawford}, Thomas M. and {Crowley}, Joseph and {David}, John and {Derome}, Mark and {Dexter}, Matthew and {Dornbusch}, Sven and {Dudevoir}, Kevin A. and {Dzib}, Sergio A. and {Eckart}, Andreas and {Eckert}, Chris and {Erickson}, Neal R. and {Everett}, Wendeline B. and {Faber}, Aaron and {Farah}, Joseph R. and {Fath}, Vernon and {Folkers}, Thomas W. and {Forbes}, David C. and {Freund}, Robert and {G{\'o}mez-Ruiz}, Arturo I. and {Gale}, David M. and {Gao}, Feng and {Geertsema}, Gertie and {Graham}, David A. and {Greer}, Christopher H. and {Grosslein}, Ronald and {Gueth}, Fr{\'e}d{\'e}ric and {Haggard}, Daryl and {Halverson}, Nils W. and {Han}, Chih-Chiang and {Han}, Kuo-Chang and {Hao}, Jinchi and {Hasegawa}, Yutaka and {Henning}, Jason W. and {Hern{\'a}ndez-G{\'o}mez}, Antonio and {Herrero-Illana}, Rub{\'e}n and {Heyminck}, Stefan and {Hirota}, Akihiko and {Hoge}, James and {Huang}, Yau-De and {Impellizzeri}, C.~M. Violette and {Jiang}, Homin and {Kamble}, Atish and {Keisler}, Ryan and {Kimura}, Kimihiro and {Kono}, Yusuke and {Kubo}, Derek and {Kuroda}, John and {Lacasse}, Richard and {Laing}, Robert A. and {Leitch}, Erik M. and {Li}, Chao-Te and {Lin}, Lupin C. -C. and {Liu}, Ching-Tang and {Liu}, Kuan-Yu and {Lu}, Li-Ming and {Marson}, Ralph G. and {Martin-Cocher}, Pierre L. and {Massingill}, Kyle D. and {Matulonis}, Callie and {McColl}, Martin P. and {McWhirter}, Stephen R. and {Messias}, Hugo and {Meyer-Zhao}, Zheng and {Michalik}, Daniel and {Monta{\~n}a}, Alfredo and {Montgomerie}, William and {Mora-Klein}, Matias and {Muders}, Dirk and {Nadolski}, Andrew and {Navarro}, Santiago and {Neilsen}, Joseph and {Nguyen}, Chi H. and {Nishioka}, Hiroaki and {Norton}, Timothy and {Nowak}, Michael A. and {Nystrom}, George and {Ogawa}, Hideo and {Oshiro}, Peter and {Oyama}, Tomoaki and {Parsons}, Harriet and {Paine}, Scott N. and {Pe{\~n}alver}, Juan and {Phillips}, Neil M. and {Poirier}, Michael and {Pradel}, Nicolas and {Primiani}, Rurik A. and {Raffin}, Philippe A. and {Rahlin}, Alexandra S. and {Reiland}, George and {Risacher}, Christopher and {Ruiz}, Ignacio and {S{\'a}ez-Mada{\'\i}n}, Alejandro F. and {Sassella}, Remi and {Schellart}, Pim and {Shaw}, Paul and {Silva}, Kevin M. and {Shiokawa}, Hotaka and {Smith}, David R. and {Snow}, William and {Souccar}, Kamal and {Sousa}, Don and {Sridharan}, T.~K. and {Srinivasan}, Ranjani and {Stahm}, William and {Stark}, Anthony A. and {Story}, Kyle and {Timmer}, Sjoerd T. and {Vertatschitsch}, Laura and {Walther}, Craig and {Wei}, Ta-Shun and {Whitehorn}, Nathan and {Whitney}, Alan R. and {Woody}, David P. and {Wouterloot}, Jan G.~A. and {Wright}, Melvin and {Yamaguchi}, Paul and {Yu}, Chen-Yu and {Zeballos}, Milagros and {Zhang}, Shuo and {Ziurys}, Lucy},
        title = "{First M87 Event Horizon Telescope Results. I. The Shadow of the Supermassive Black Hole}",
      journal = {\apjl},
         year = 2019,
        month = apr,
       volume = {875},
       number = {1},
          eid = {L1},
        pages = {L1},
          doi = {10.3847/2041-8213/ab0ec7},
archivePrefix = {arXiv},
       eprint = {1906.11238},
 primaryClass = {astro-ph.GA},
       adsurl = {https://ui.adsabs.harvard.edu/abs/2019ApJ...875L...1E}
}

@ARTICLE{EHT21a,
       author = {{EHTC} and {Akiyama}, Kazunori and {Algaba}, Juan Carlos and {Alberdi}, Antxon and {Alef}, Walter and {Anantua}, Richard and {Asada}, Keiichi and {Azulay}, Rebecca and {Baczko}, Anne-Kathrin and {Ball}, David},
        title = "{First M87 Event Horizon Telescope Results. VII. Polarization of the Ring}",
      journal = {\apjl},
         year = 2021,
        month = mar,
       volume = {910},
       number = {1},
          eid = {L12},
        pages = {L12},
          doi = {10.3847/2041-8213/abe71d},
archivePrefix = {arXiv},
       eprint = {2105.01169},
 primaryClass = {astro-ph.HE},
       adsurl = {https://ui.adsabs.harvard.edu/abs/2021ApJ...910L..12E}
}

@ARTICLE{EHT22a,
       author = {{EHTC} and {Akiyama}, Kazunori and {Alberdi}, Antxon and {Alef}, Walter and {Algaba}, Juan Carlos and {Anantua}, Richard and {Asada}, Keiichi and {Azulay}, Rebecca and {Bach}, Uwe and {Baczko}, Anne-Kathrin and {Ball}, David and {Balokovi{\'c}}, Mislav and {Barrett}, John and {Baub{\"o}ck}, Michi and {Benson}, Bradford A. and {Bintley}, Dan and {Blackburn}, Lindy and {Blundell}, Raymond and {Bouman}, Katherine L. and {Bower}, Geoffrey C. and {Boyce}, Hope and {Bremer}, Michael and {Brinkerink}, Christiaan D. and {Brissenden}, Roger and {Britzen}, Silke and {Broderick}, Avery E. and {Broguiere}, Dominique and {Bronzwaer}, Thomas and {Bustamante}, Sandra and {Byun}, Do-Young and {Carlstrom}, John E. and {Ceccobello}, Chiara and {Chael}, Andrew and {Chan}, Chi-kwan and {Chatterjee}, Koushik and {Chatterjee}, Shami and {Chen}, Ming-Tang and {Chen}, Yongjun and {Cheng}, Xiaopeng and {Cho}, Ilje and {Christian}, Pierre and {Conroy}, Nicholas S. and {Conway}, John E. and {Cordes}, James M. and {Crawford}, Thomas M. and {Crew}, Geoffrey B. and {Cruz-Osorio}, Alejandro and {Cui}, Yuzhu and {Davelaar}, Jordy and {De Laurentis}, Mariafelicia and {Deane}, Roger and {Dempsey}, Jessica and {Desvignes}, Gregory and {Dexter}, Jason and {Dhruv}, Vedant and {Doeleman}, Sheperd S. and {Dougal}, Sean and {Dzib}, Sergio A. and {Eatough}, Ralph P. and {Emami}, Razieh and {Falcke}, Heino and {Farah}, Joseph and {Fish}, Vincent L. and {Fomalont}, Ed and {Ford}, H. Alyson and {Fraga-Encinas}, Raquel and {Freeman}, William T. and {Friberg}, Per and {Fromm}, Christian M. and {Fuentes}, Antonio and {Galison}, Peter and {Gammie}, Charles F. and {Garc{\'\i}a}, Roberto and {Gentaz}, Olivier and {Georgiev}, Boris and {Goddi}, Ciriaco and {Gold}, Roman and {G{\'o}mez-Ruiz}, Arturo I. and {G{\'o}mez}, Jos{\'e} L. and {Gu}, Minfeng and {Gurwell}, Mark and {Hada}, Kazuhiro and {Haggard}, Daryl and {Haworth}, Kari and {Hecht}, Michael H. and {Hesper}, Ronald and {Heumann}, Dirk and {Ho}, Luis C. and {Ho}, Paul and {Honma}, Mareki and {Huang}, Chih-Wei L. and {Huang}, Lei and {Hughes}, David H. and {Ikeda}, Shiro and {Impellizzeri}, C.~M. Violette and {Inoue}, Makoto and {Issaoun}, Sara and {James}, David J. and {Jannuzi}, Buell T. and {Janssen}, Michael and {Jeter}, Britton and {Jiang}, Wu and {Jim{\'e}nez-Rosales}, Alejandra and {Johnson}, Michael D. and {Jorstad}, Svetlana and {Joshi}, Abhishek V. and {Jung}, Taehyun and {Karami}, Mansour and {Karuppusamy}, Ramesh and {Kawashima}, Tomohisa and {Keating}, Garrett K. and {Kettenis}, Mark and {Kim}, Dong-Jin and {Kim}, Jae-Young and {Kim}, Jongsoo and {Kim}, Junhan and {Kino}, Motoki and {Koay}, Jun Yi and {Kocherlakota}, Prashant and {Kofuji}, Yutaro and {Koch}, Patrick M. and {Koyama}, Shoko and {Kramer}, Carsten and {Kramer}, Michael and {Krichbaum}, Thomas P. and {Kuo}, Cheng-Yu and {La Bella}, Noemi and {Lauer}, Tod R. and {Lee}, Daeyoung and {Lee}, Sang-Sung and {Leung}, Po Kin and {Levis}, Aviad and {Li}, Zhiyuan and {Lico}, Rocco and {Lindahl}, Greg and {Lindqvist}, Michael and {Lisakov}, Mikhail and {Liu}, Jun and {Liu}, Kuo and {Liuzzo}, Elisabetta and {Lo}, Wen-Ping and {Lobanov}, Andrei P. and {Loinard}, Laurent and {Lonsdale}, Colin J. and {Lu}, Ru-Sen and {Mao}, Jirong and {Marchili}, Nicola and {Markoff}, Sera and {Marrone}, Daniel P. and {Marscher}, Alan P. and {Mart{\'\i}-Vidal}, Iv{\'a}n and {Matsushita}, Satoki and {Matthews}, Lynn D. and {Medeiros}, Lia and {Menten}, Karl M. and {Michalik}, Daniel and {Mizuno}, Izumi and {Mizuno}, Yosuke and {Moran}, James M. and {Moriyama}, Kotaro and {Moscibrodzka}, Monika and {M{\"u}ller}, Cornelia and {Mus}, Alejandro and {Musoke}, Gibwa and {Myserlis}, Ioannis and {Nadolski}, Andrew and {Nagai}, Hiroshi and {Nagar}, Neil M. and {Nakamura}, Masanori and {Narayan}, Ramesh and {Narayanan}, Gopal and {Natarajan}, Iniyan and {Nathanail}, Antonios and {Navarro Fuentes}, Santiago and {Neilsen}, Joey and {Neri}, Roberto and {Ni}, Chunchong and {Noutsos}, Aristeidis and {Nowak}, Michael A. and {Oh}, Junghwan and {Okino}, Hiroki and {Olivares}, H{\'e}ctor and {Ortiz-Le{\'o}n}, Gisela N. and {Oyama}, Tomoaki and {{\"O}zel}, Feryal and {Palumbo}, Daniel C.~M. and {Paraschos}, Georgios Filippos and {Park}, Jongho and {Parsons}, Harriet and {Patel}, Nimesh and {Pen}, Ue-Li and {Pesce}, Dominic W. and {Pi{\'e}tu}, Vincent and {Plambeck}, Richard and {PopStefanija}, Aleksandar and {Porth}, Oliver and {P{\"o}tzl}, Felix M. and {Prather}, Ben and {Preciado-L{\'o}pez}, Jorge A. and {Psaltis}, Dimitrios and {Pu}, Hung-Yi and {Ramakrishnan}, Venkatessh and {Rao}, Ramprasad and {Rawlings}, Mark G. and {Raymond}, Alexander W. and {Rezzolla}, Luciano and {Ricarte}, Angelo and {Ripperda}, Bart and {Roelofs}, Freek and {Rogers}, Alan and {Ros}, Eduardo and {Romero-Ca{\~n}izales}, Cristina and {Roshanineshat}, Arash and {Rottmann}, Helge and {Roy}, Alan L. and {Ruiz}, Ignacio and {Ruszczyk}, Chet and {Rygl}, Kazi L.~J. and {S{\'a}nchez}, Salvador and {S{\'a}nchez-Arg{\"u}elles}, David and {S{\'a}nchez-Portal}, Miguel and {Sasada}, Mahito and {Satapathy}, Kaushik and {Savolainen}, Tuomas and {Schloerb}, F. Peter and {Schonfeld}, Jonathan and {Schuster}, Karl-Friedrich and {Shao}, Lijing and {Shen}, Zhiqiang and {Small}, Des and {Sohn}, Bong Won and {SooHoo}, Jason and {Souccar}, Kamal and {Sun}, He and {Tazaki}, Fumie and {Tetarenko}, Alexandra J. and {Tiede}, Paul and {Tilanus}, Remo P.~J. and {Titus}, Michael and {Torne}, Pablo and {Traianou}, Efthalia and {Trent}, Tyler and {Trippe}, Sascha and {Turk}, Matthew and {van Bemmel}, Ilse and {van Langevelde}, Huib Jan and {van Rossum}, Daniel R. and {Vos}, Jesse and {Wagner}, Jan and {Ward-Thompson}, Derek and {Wardle}, John and {Weintroub}, Jonathan and {Wex}, Norbert and {Wharton}, Robert and {Wielgus}, Maciek and {Wiik}, Kaj and {Witzel}, Gunther and {Wondrak}, Michael F. and {Wong}, George N. and {Wu}, Qingwen and {Yamaguchi}, Paul and {Yoon}, Doosoo and {Young}, Andr{\'e} and {Young}, Ken and {Younsi}, Ziri and {Yuan}, Feng and {Yuan}, Ye-Fei and {Zensus}, J. Anton and {Zhang}, Shuo and {Zhao}, Guang-Yao and {Zhao}, Shan-Shan and {Agurto}, Claudio and {Allardi}, Alexander and {Amestica}, Rodrigo and {Araneda}, Juan Pablo and {Arriagada}, Oriel and {Berghuis}, Jennie L. and {Bertarini}, Alessandra and {Berthold}, Ryan and {Blanchard}, Jay and {Brown}, Ken and {C{\'a}rdenas}, Mauricio and {Cantzler}, Michael and {Caro}, Patricio and {Castillo-Dom{\'\i}nguez}, Edgar and {Chan}, Tin Lok and {Chang}, Chih-Cheng and {Chang}, Dominic O. and {Chang}, Shu-Hao and {Chang}, Song-Chu and {Chen}, Chung-Chen and {Chilson}, Ryan and {Chuter}, Tim C. and {Ciechanowicz}, Miroslaw and {Colin-Beltran}, Edgar and {Coulson}, Iain M. and {Crowley}, Joseph and {Degenaar}, Nathalie and {Dornbusch}, Sven and {Dur{\'a}n}, Carlos A. and {Everett}, Wendeline B. and {Faber}, Aaron and {Forster}, Karl and {Fuchs}, Miriam M. and {Gale}, David M. and {Geertsema}, Gertie and {Gonz{\'a}lez}, Edouard and {Graham}, Dave and {Gueth}, Fr{\'e}d{\'e}ric and {Halverson}, Nils W. and {Han}, Chih-Chiang and {Han}, Kuo-Chang and {Hasegawa}, Yutaka and {Hern{\'a}ndez-Rebollar}, Jos{\'e} Luis and {Herrera}, Cristian and {Herrero-Illana}, Ruben and {Heyminck}, Stefan and {Hirota}, Akihiko and {Hoge}, James and {Hostler Schimpf}, Shelbi R. and {Howie}, Ryan E. and {Huang}, Yau-De and {Jiang}, Homin and {Jinchi}, Hao and {John}, David and {Kimura}, Kimihiro and {Klein}, Thomas and {Kubo}, Derek and {Kuroda}, John and {Kwon}, Caleb and {Lacasse}, Richard and {Laing}, Robert and {Leitch}, Erik M. and {Li}, Chao-Te and {Liu}, Ching-Tang and {Liu}, Kuan-Yu and {Lin}, Lupin C. -C. and {Lu}, Li-Ming and {Mac-Auliffe}, Felipe and {Martin-Cocher}, Pierre and {Matulonis}, Callie and {Maute}, John K. and {Messias}, Hugo and {Meyer-Zhao}, Zheng and {Monta{\~n}a}, Alfredo and {Montenegro-Montes}, Francisco and {Montgomerie}, William and {Moreno Nolasco}, Marcos Emir and {Muders}, Dirk and {Nishioka}, Hiroaki and {Norton}, Timothy J. and {Nystrom}, George and {Ogawa}, Hideo and {Olivares}, Rodrigo and {Oshiro}, Peter and {P{\'e}rez-Beaupuits}, Juan Pablo and {Parra}, Rodrigo and {Phillips}, Neil M. and {Poirier}, Michael and {Pradel}, Nicolas and {Qiu}, Richard and {Raffin}, Philippe A. and {Rahlin}, Alexandra S. and {Ram{\'\i}rez}, Jorge and {Ressler}, Sean and {Reynolds}, Mark and {Rodr{\'\i}guez-Montoya}, Iv{\'a}n and {Saez-Madain}, Alejandro F. and {Santana}, Jorge and {Shaw}, Paul and {Shirkey}, Leslie E. and {Silva}, Kevin M. and {Snow}, William and {Sousa}, Don and {Sridharan}, T.~K. and {Stahm}, William and {Stark}, Anthony A. and {Test}, John and {Torstensson}, Karl and {Venegas}, Paulina and {Walther}, Craig and {Wei}, Ta-Shun and {White}, Chris and {Wieching}, Gundolf and {Wijnands}, Rudy and {Wouterloot}, Jan G.~A. and {Yu}, Chen-Yu and {Yu}, Wei and {Zeballos}, Milagros and {EHT Collaboration}},
        title = "{First Sagittarius A* Event Horizon Telescope Results. I. The Shadow of the Supermassive Black Hole in the Center of the Milky Way}",
      journal = {\apjl},
         year = 2022,
        month = may,
       volume = {930},
       number = {2},
          eid = {L12},
        pages = {L12},
          doi = {10.3847/2041-8213/ac6674},
       adsurl = {https://ui.adsabs.harvard.edu/abs/2022ApJ...930L..12A}
}

@ARTICLE{Fromm16,
       author = {{Fromm}, C.~M. and {Perucho}, M. and {Mimica}, P. and {Ros}, E.},
        title = "{Spectral evolution of flaring blazars from numerical simulations}",
      journal = {\aap},
         year = 2016,
        month = apr,
       volume = {588},
          eid = {A101},
        pages = {A101},
          doi = {10.1051/0004-6361/201527139},
archivePrefix = {arXiv},
       eprint = {1601.03181},
 primaryClass = {astro-ph.HE},
       adsurl = {https://ui.adsabs.harvard.edu/abs/2016A&A...588A.101F}
}

@ARTICLE{Guarnieri25,
       author = {{Monti-Guarnieri}, Pietro and {Bernard}, Denis},
        title = "{Fermi-LAT detection of renewed gamma-ray activity from the radio galaxy NGC 1275}",
      journal = {The Astronomer's Telegram},
         year = 2025,
        month = jan,
       volume = {16988},
        pages = {1},
       adsurl = {https://ui.adsabs.harvard.edu/abs/2025ATel16988....1M}
}

\clearpage 
\begin{appendix}

\begin{figure*}[t!] 
    \centering
    
    \begin{minipage}{\textwidth}
        \section{\texttt{CLEAN} reconstruction of 43\,GHz data}\label{app:BU}
        \centering
        \includegraphics[scale=0.52]{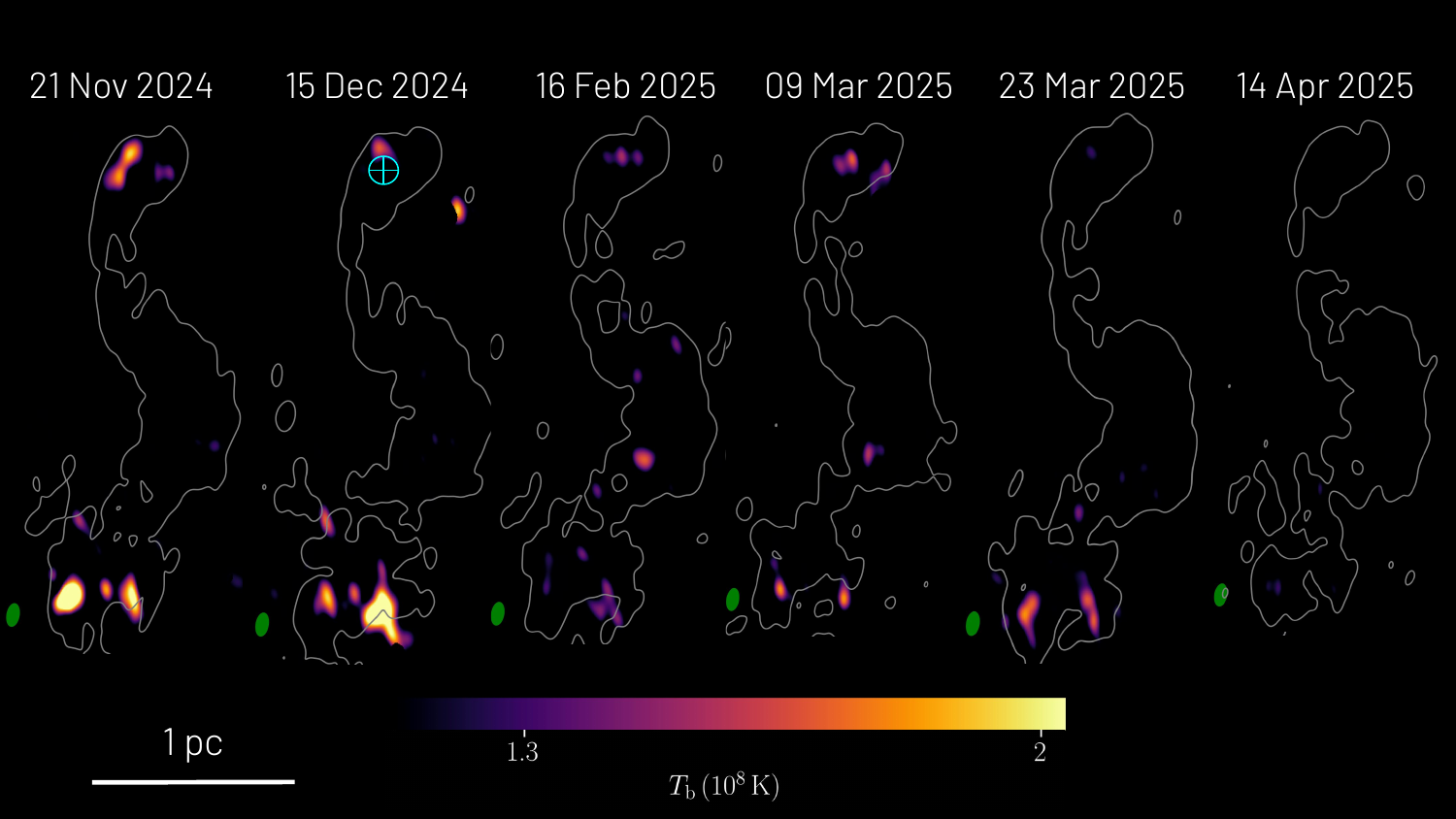} 
        \captionof{figure}{
            Publicly available \texttt{CLEAN}-reconstruction images of \C\ at 43\,GHz (BEAM-ME).
            The display is in a similar manner to the bottom panel of Fig.~\ref{fig:BU}.
            The green ellipse next to each reconstruction illustrates the restoring \textrm{CLEAN}-beam, which corresponds to $(0.15\times0.30)\,\textrm{mas}\,(10^\circ)$ on average.
            We note the consistency between the RML and \texttt{CLEAN} reconstructions.
            Specifically, the core region of \C\ exhibits a lower linear polarisation signal, whereas the C3 region appears brighter in the November and December 2024 epochs (around the time of the flare) before also returning to a lower signal state.
        }
        \label{fig:BU2}
    \end{minipage}

    \vspace{3em} 

        \begin{minipage}{\textwidth}
        \section{Total intensity flux density light curve}\label{app:LC}
        \centering
        \includegraphics[scale=0.42]{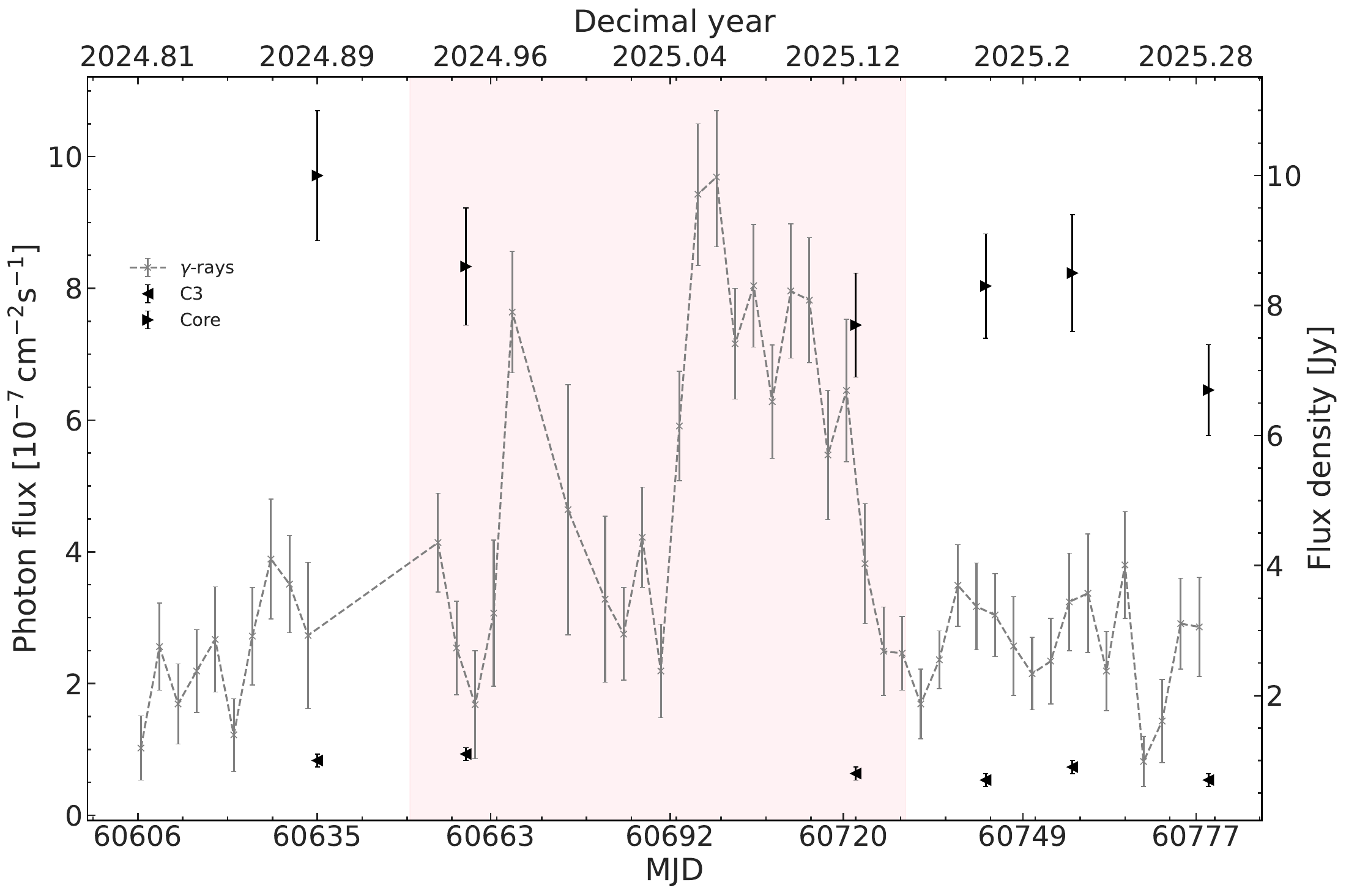}
        \captionof{figure}{
            Total intensity flux density light curve of the core and C3 region, displayed along the $\gamma$-ray light curve.
            The setup of the figure is similar to Fig.~\ref{fig:SingleDish}.
            While the core flux density remains stable during the $\gamma$-ray flare, the one of the C3 region is higher, following the same trend as the polarisation degree.
        }
        \label{fig:SingleDishSI}
    \end{minipage}
\end{figure*}

\end{appendix}

\end{document}